\def\year{2024}\relax
\PassOptionsToPackage{table}{xcolor}
\documentclass[letterpaper]{article} %
\usepackage{aaai24}  %
\usepackage{times}  %
\usepackage{helvet}  %
\usepackage{courier}  %
\usepackage[hyphens]{url}  %
\usepackage{graphicx} %
\usepackage{natbib}  %
\usepackage{caption} %
\usepackage{algorithm}
\usepackage{algorithmic}
\usepackage{tabularx}

\usepackage{array, makecell} %
\usepackage{multirow}

\usepackage{xspace}
\newcommand{\dsrumble}{\(D_{rumble}\)\xspace}
\newcommand{\dsyoutube}{\(D_{youtube}\)\xspace}
\newcommand{\dspolitical}{\(D_{political}\)\xspace}
\newcommand{\dsleft}{\(D_{left}\)\xspace}
\newcommand{\dscenter}{\(D_{center}\)\xspace}
\newcommand{\dsright}{\(D_{right}\)\xspace}

\usepackage{xcolor}
\usepackage{newfloat}
\usepackage{listings}
\DeclareCaptionStyle{ruled}{labelfont=normalfont,labelsep=colon,strut=off} %
\floatstyle{ruled}
\newfloat{listing}{tb}{lst}{}
\floatname{listing}{Listing}
\usepackage{booktabs}
\usepackage{amssymb} 
\usepackage{graphicx}
\usepackage{subcaption}
\usepackage[table]{xcolor}

\newcommand{\descr}[1]{\smallskip\noindent\textbf{#1}}

\title{Podcast Outcasts: Understanding Rumble's Podcast Dynamics }

\usepackage{bibentry}

\author {
    Utkucan Balci\textsuperscript{\rm 1},
    Jay Patel \textsuperscript{\rm 1},
    Berkan Balci \textsuperscript{\rm 2},
    Jeremy Blackburn \textsuperscript{\rm 1}
}
\affiliations {
    \textsuperscript{\rm 1} Binghamton University\\
    \textsuperscript{\rm 2} Middle East Technical University	\\
    ubalci1@binghamton.edu, jpatel67@binghamton.edu, 
    e252615@metu.edu.tr, jblackbu@binghamton.edu
}

\begin{document}

\maketitle

\begin{abstract}
  Podcasting on Rumble, an alternative video-sharing platform, attracts controversial figures known for spreading divisive and often misleading content, which sharply contrasts with YouTube's more regulated environment.
  Motivated by the growing impact of podcasts on political discourse, as seen with figures like Joe Rogan and Andrew Tate, this paper explores the political biases and content strategies used by these platforms.
  In this paper, we conduct a comprehensive analysis of over 13K podcast videos from both YouTube and Rumble, focusing on their political content and the dynamics of their audiences.
  Using advanced speech-to-text transcription, topic modeling, and contrastive learning techniques, we explore three critical aspects: the presence of political bias in podcast channels, the nature of content that drives podcast views, and the usage of visual elements in these podcasts.
  Our findings reveal a distinct right-wing orientation in Rumble's podcasts, contrasting with YouTube's more diverse and apolitical content. 

\end{abstract}

\section{Introduction}
\label{sec:intro}

In today's world, visual elements play an important role in communication and engagement~\cite{ling2021dissecting}.
The rise of social media, video-sharing platforms, and visual-centric content has transformed how we perceive information.
This shift has transformed various media formats, including podcasts.
The integration of visual elements into podcasts has given rise to video podcasts, making it increasingly popular~\cite{Riversidefm23}.
Ultimately, this trend has not only boosted the popularity of the video podcasts but also led YouTube to match Spotify's reach in 2022 and then surpass it as the most accessed platform for podcasts in 2023, which in turn has pushed Spotify and Apple to broaden their video podcast services~\cite{Mayer2023b}.

A notable illustration of this impact is Joe Rogan, a controversial personality known for several scandalous incidents, including the use of racial slurs in an Instagram clip~\cite{MarufStelter}.
In 2022, he accepted a \$200M offer from Spotify for three and a half years, moving his content exclusively from YouTube to Spotify~\cite{Rosman2022}. 
Rogan also declined a \$100M, four year offer from Rumble, an alternative video sharing platform~\cite{Spangler2022}.
Another contentious figure, Andrew Tate, has been banned from Facebook, Instagram, TikTok, and YouTube for his misogynistic rhetoric~\cite{Wilson2022}.
As of early 2024, he continues to share his content on Rumble, where he has surpassed 1.7 million followers~\cite{TateSpeechRumble}.

The list of controversial political figures associated with Rumble extends beyond Andrew Tate. 
By 2024, Rumble is a platform for numerous individuals who have encountered restrictions on YouTube, including former President Trump~\cite{Mak2021}, conspiracy theorist Alex Jones~\cite{Farah2023}, and comedian Russell Brand~\cite{Klee2023}.
Despite this YouTube to Rumble pipeline of controversial right-wing figures, Rumble's owner claims the platform is ``neutral~\cite{Brown2022},'' and more than one third of its user base is Democrat~\cite{Rainey2023}.
To date, no research has explored whether right-wing podcasters are merely a segment of Rumble's podcasting selection or if the platform serves as a bastion for right-wing propaganda.

In this paper, we analyze over 13K podcast videos, equivalent to 526 days of content, aiming to understand the focus of popular YouTube and Rumble podcast channels, particularly in terms of their political content.
Our investigation is centered around three key research questions:

\descr{RQ1:}  Is there a political bias in the podcast videos on Rumble?

\descr{RQ2:} What kind of content drives user views on podcast videos of popular Rumble channels, and how does this compare to YouTube podcast channels?

\descr{RQ3:} What visual elements are most prevalent in popular podcast channels on Rumble and YouTube, and how do these elements align with the political leanings of the content?

First, we analyze speech-to-text transcriptions generated by a modification of OpenAI's Whisper model, faster-whisper, to understand whether there is a detectable political bias in the content of popular podcasts on Rumble compared to those on YouTube. 
This involves an examination of the themes, narrative structures, and ideological tendencies present in these podcast videos.
For this purpose, we apply a topic modeling approach that uses transformer-based neural network embeddings. 
This method leverages contextual embeddings to understand and categorize the content of the podcasts more accurately.

Second, we investigate the types of discussions that drive user engagement on podcast videos of these channels. 
We examine the link between the characteristics of content and audience involvement, aiming to reveal how content popularity can impact political discourse.

In our final analysis, we investigate the use of visual elements in these podcast videos through embeddings generated by contrastive learning vision model.
The existing body of research on visual elements in podcasts is limited, focusing mainly on their use in educational contexts and noting a positive impact on student engagement~\cite{putri2022radio,rajic2013educational}.
Considering that visual elements can positively influence virality, particularly through the presence of characters and their poses~\cite{ling2021dissecting}, we aim to explore how popular podcast channels on Rumble and YouTube incorporate visual elements in their content.

Our research reveals a clear pattern: Rumble exhibits a noticeable right-wing bias in its audio and visual content, whereas YouTube primarily remains apolitical, concentrating on mainstream subjects.
Furthermore, our study indicates that the cancellation of a prominent right-wing personality and featuring controversial content related to COVID-19 significantly impacts Rumble's podcast views.

\section{Background \& Related Work}
\label{sec:background-related-work}

The term ``podcast'' was mentioned for the first time in 2004~\cite{Robertson2019}.
In 2006, the PEW Research Center provided a definition~\cite{Pew2006} as follows:
``\textit{Podcasting is a way to distribute audio and video programming over the Web that differs from earlier online audio and video publishing because the material is automatically transferred to the user’s computer and can be consumed at any time, usually on an Apple iPod or another kind of portable digital music player commonly known as an MP3 player.}''
Although the core of this definition remains relevant, the influence of podcasts has evolved remarkably over time. 
Particularly with the widespread use of social media, podcasts are now viewed by millions of users on video streaming platforms~\cite{Escandon2024b}.

In a 12-month period spanning parts of both 2022 and 2023, nearly half of the US adult population reported having listened to a podcast, with one-fifth frequently doing so multiple times a week~\cite{Pew2023}.
This proportion increases to one-third among young adults under 30. 
Of the U.S. adults who listened to a podcast during this interval, 46\% were Republicans and 54\% Democrats, with 65\% of Republican and 69\% of Democratic listeners tuning into news-related podcasts.

\descr{Podcasts as vectors of political discourse.}
A range of studies have explored the impact of political podcasts on individuals' political engagement and attitudes~\cite{choMotivesUsingNews2023, eurittPublicCirculationNPR2019, leeNewsPodcastUsage2021a, kimDelineatingComplexUse2016, macdougallPodcastingPoliticalLife2011,raePodcastsPoliticalListening2023,sternePoliticsPodcasting2008}.
Notably, consuming podcasts is linked to heightened levels of personalized politics, a process where individuals integrate new information into their existing ideological frameworks to develop more personalized political understandings~\cite{bratcherDeeperDiscussionSurvey2022a}. 
\cite{kimSelectiveExposurePodcast2016} further explored the relationship between partisan podcast consumption, emotional responses, and political participation, finding that selective exposure to partisan podcasts can shape emotional reactions to political candidates, thereby affecting political engagement. 
\cite{chadhaListeningBuildingProfile2012} also observed a positive correlation between using podcasts for news and increased political participation, suggesting that podcasts might boost political involvement among individuals. 
While many of these studies have focused on YouTube podcasts and estimating the ideology of YouTube channels~\cite{dinkovPredictingLeadingPolitical2019,lai2022estimating}, there has yet to be a large-scale, data-driven analysis of the political bias in popular YouTube podcast channels.

\descr{What is Rumble?}
Launched in 2013 as a YouTube alternative, Rumble gained notable attention during the COVID-19 pandemic~\cite{McCluskey2022}. 
The number of monthly users on the platform increased from 1.6 million in the Fall 2020 to 31.9 million by the beginning of 2021~\cite{Pramod2021} and eventually hit a peak of 80 million active users monthly by the end of 2022~\cite{Brown2022}.
While the platform's founder asserts its neutrality~\cite{Brown2022}, Rumble has become particularly known for being a haven for right-leaning public figures, including Andrew Tate, Rudy Giuliani, and Alex Jones~\cite{Farah2023}. 
Despite its popularity, research on this platform is limited.
Previous work~\cite{stocking2022role} estimated that over 75\% of US adults who regularly use Rumble for news are Republicans or lean towards the Republican Party.
This survey also notes that Rumble is a regular news source for 2\% of the American population.
While Rumble has been mentioned in research related to the alt-right and the Russian invasion of Ukraine~\cite{chen2023tweets,aliapoulios2021early,aliapoulios2021large}, and some of Andrew Tate's Rumble channel podcast episodes have undergone analysis~\cite{sayogie2023patriarchal}, similar to YouTube, a large-scale, data-driven research analyzing political bias in popular Rumble podcast channels has yet to be conducted.

\begin{table*}[th!]	
\centering
\scriptsize
  \begin{tabular}{lrrr|lrrr}
  \toprule
  \multicolumn{4}{c|}{YouTube} & \multicolumn{4}{c}{Rumble} \\
  \midrule
Channel &  \# Views &  \# Podcasts & Avg. Views & Channel &  \# Views &  \# Podcasts & Avg. Views \\

\midrule
H3 Podcast                              & 183M & 108 & 1.7M & The Dan Bongino Show      &    133M &          576 &        231K \\

Philip DeFranco                         & 143M & 156 & 918K & Steven Crowder            &     42M &          212 &        198K \\

rSlash                                  & 111M & 223 & 500K & The Post Millennial       &     13M &           10 &        1.3M \\

No Jumper                               & 107M & 465 & 231K & RepMattGaetz              &      9.7M &           45 &        216K \\

Bailey Sarian                           & 10M & 34  & 3M    & TateSpeech by Andrew Tate &      7.8M &            3 &        2.6M \\

IMPAULSIVE                              & 89M  & 39  & 2M   & The JD Rucker Show        &      7.3M &           38 &        194K \\

REVOLT                                  & 88M  & 61  & 1.4M & The Charlie Kirk Show     &      5.7M &          215 &         26K \\

YMH Studios                             & 77M  & 130 & 598K & The Rubin Report          &      5.0M &          174 &         28K \\

Gecko's Garage - Trucks For Children    & 70M  & 42  & 1.6M & Glenn Greenwald           &      4.8M &           24 &        201K \\

FLAGRANT                                & 67M  & 51  & 1.3M & HodgeTwins                &      4.6M &          152 &         30K \\

Dr. Sten Ekberg                         & 64M  & 50  & 1.3M & Senator Ron Johnson       &      4.5M &            1 &        4.5M \\

Lex Fridman                             & 64M  & 59  & 1M   & Devin Nunes               &      4.2M &           64 &         66K \\

The 85 South Comedy Show                & 63M  & 45  & 1.4M & vivafrei                  &      4.2M &          178 &         23K \\

NBC News                                & 61M  & 313 & 196K & Dinesh D'Souza            &      4.1M &          208 &         20K \\

The Pat McAfee Show                     & 58M  & 161 & 365K & Russell Brand             &      4.0M &           48 &         83K \\

FreshandFit                             & 55M  & 226 & 246K & TheSaltyCracker           &      3.8M &           62 &         62K \\

Critical Role                           & 51M  & 26  & 1.9M & Ben Shapiro               &      3.2M &          297 &         10K \\

CinnamonToastKen                        & 47M  & 41  & 1.1M & TimcastIRL                &      3.1M &          326 &          9K \\

Jordan B Peterson                       & 47M  & 43  & 1M   & The Trish Regan Show      &      2.9M &          190 &         15K \\

48 Hours	                              & 46M  & 10  & 4.6M & Joe Pags                  &      2.4M &          134 &         18K \\
\bottomrule
\end{tabular}
\caption{Top 20 podcast video channels of YouTube and Rumble, by their cumulative views, total number of videos, and average views.}
\label{tab:YouTube_rumble_top_20_by_view}
\end{table*}

\section{Dataset}
\label{sec:dataset}

To collect podcast videos from Rumble, we develop a custom crawler that extracts video information from the ``Podcasts'' section on the home page of rumble.com~\cite{rumblepodcasts}.
This crawler systematically navigates through the URLs, scanning pages in this section until no new pages are found. 
We initially ran our crawler in October 2022 and conducted a follow-up in early 2023 to ensure coverage of the entire year.
In the first week of July 2023, we revisited the video pages in our collection to update their metadata and remove any podcast videos that were no longer accessible.
The rationale for this approach is to allow at least six full months for the metadata of each video (e.g., views) to stabilize and reflect their actual values.
As a result, we compile a dataset of 6,761 videos from 246 channels, posted between August 27, 2020, to January 1, 2023.
To remove non-English content from our dataset we perform language verification and transcribe the podcast videos using a reimplementation of OpenAI's Whisper~\cite{OpenAIWhisper}.
Further details of these steps can be found in Appendix A. 
A more comprehensive look at this dataset can be found in the corresponding dataset paper~\cite{balci2024idrama}.
As we aim to analyze popular Rumble podcast channels, we limit our dataset to include the top 100 channels with the highest cumulative podcast video views.
This subset comprises a total of 6,272 videos, accounting for 99\% of all podcast views on Rumble.
Table~\ref{tab:YouTube_rumble_top_20_by_view} presents the top 20 channels by cumulative views, along with their total number of videos and average view counts in our dataset.
We refer to this dataset as \dsrumble throughout the remainder of this paper.

\descr{YouTube.}
Using the YouTube API, we extract video metadata categorized as podcasts from YouTube's list of top 100 popular podcast creators~\cite{YouTubepopular}.
Our manual inspection of these channels revealed non-English content and videos unrelated to podcasts (e.g., music and gospel).
To refine our dataset, we implemented the following criteria:
1)~videos must be categorized under the Podcast tab within the channel's playlists,
2)~the content must be in English, and
3)~genres unrelated to podcasts (e.g., gospel and music) are excluded. 
In the refinement process, we randomly select and manually inspect 5 videos from each playlist, subsequently eliminating playlists that failed to meet our criteria.
This process yields a dataset of more than 20K videos from 69 channels, with all videos available and their metadata collected during the first week of July 2023.
For a comparative analysis with the Rumble dataset, we adjust the YouTube dataset to match the monthly video distribution and the total number of podcast videos in the Rumble dataset.
This way, by aligning the dataset with the specific months, we account for the potential influence of simultaneous events on the focus and content of the discussions.
Next, we eliminate non-English content and transcribed the podcast videos following the methodologies outlined in Appendix A.
Overall, we collect 6,272 podcast videos using youtube-dl~\cite{youtubedl}.
Table~\ref{tab:YouTube_rumble_top_20_by_view} displays the top 20 channels by cumulative views, including their total number of videos and average view counts in our YouTube dataset.
We refer to this dataset as \dsyoutube throughout the remainder of this paper.

\descr{Political podcast channels.}
To compare \dsrumble and \dsyoutube politically, we adopt a pre-established~\cite{ dinkovPredictingLeadingPolitical2019} classification of YouTube channels into ideological categories: left, center, and right.
Previous work used this list to compute the ideological social dimensions of YouTube channels~\cite{boesinger2023tube2vec}.
For this subset of channels, we applied the same selection and refinement process used in our \dsyoutube extraction.
This resulted in a collection of 7,755 videos, categorized as 1,660 Center, 3,510 Left, and 2,585 Right. 
Next, we exclude channels that appear in both \dsrumble and \dsyoutube to prevent the influence of duplicate podcasts in our analyses.
This step is crucial to prevent the influence of identical podcasts from skewing our analyses.
Finally, to ensure balanced and comparable analyses with \dsrumble and \dsyoutube, we sample 500 videos from each political category, in line with the monthly distribution pattern observed in these datasets.
We refer to this dataset as \dspolitical moving forward. 
Additionally, the subsets of \dspolitical corresponding to the left, right, and center political categories are denoted as \dsleft, \dsright, and \dscenter, respectively.

\begin{table*}[th!]
  \centering
  \scriptsize
    \begin{tabular}{rlccc|lccc}
  \toprule
  \multicolumn{5}{c|}{YouTube} & \multicolumn{4}{c}{Rumble} \\
  \midrule
  \textbf{no.} & \textbf{Top 3 Topic Words} & \textbf{Left} & \textbf{Center} & \textbf{Right} & \textbf{Top 3 Topic Words} & \textbf{Left} & \textbf{Center} & \textbf{Right} \\
  \midrule
  1 & \small{saying, im, know}  & -- & -- & -- &  \small{vaccine, vaccinated, vaccines} & \checkmark & \checkmark & \checkmark \\
  2 & \small{bengals, nfc, raiders} & -- & -- & -- &  \small{ballots, mailin, ballot} & \checkmark & \checkmark & \checkmark \\
  3 & \small{billionaire, richest, multimillionaire}  & -- & -- & -- &  \small{ukrainians, crimea, putin} & \checkmark & \checkmark & \checkmark \\
  4 & \small{niggas, nigga, ns} & -- & -- & -- &  \small{roe, abortion, abortions} & \checkmark & \checkmark & \checkmark \\
  5 & \small{book, books, chapter}  & \checkmark & \checkmark & \checkmark &  \small{mask, masks, masking} & -- & \checkmark & \checkmark \\
  6 & \small{ukrainians, crimea, putin} & \checkmark & \checkmark & \checkmark &  \small{rumble, rumbles, rumblecom} & -- & -- & -- \\
  7 & \small{interview, interviews, interviewer} & -- & \checkmark & -- &  \small{biden, bidens, joe} & \checkmark & -- & \checkmark \\
  8 & \small{feel, antioch75, wesh} & -- & -- & -- &  \small{alito, clarence, justices} & \checkmark & \checkmark &\checkmark \\
  9 & \small{lakers, clippers, nets} & -- & -- & -- &  \small{desantis, ron, desantiss} & -- & -- & -- \\
  10 & \small{vaccine, vaccinated, vaccines}  & \checkmark & \checkmark & \checkmark &  \small{democrats, dems, republicans} & \checkmark & -- & \checkmark \\
  11 & \small{entrepreneur, entrepreneurship, entrepreneurs} & -- & -- & -- &  \small{inflation, inflationary, reduction} & \checkmark & \checkmark & \checkmark \\
  12 & \small{roe, abortion, abortions} & \checkmark & \checkmark & \checkmark &  \small{book, books, chapter} & \checkmark & \checkmark & \checkmark  \\
  13 & \small{rudn, stk, know} & -- & -- & -- &  \small{mainstream, media, medias} & -- & -- & -- \\
  14 & \footnotesize{masks, mask, masking} & -- & \checkmark & \checkmark &  \small{lefties, leftism, lefts} & -- & -- & \checkmark \\
  15 & \small{rapping, rap, hiphop}  & -- & -- & -- &  \small{tweet, retweeted, retweet} & -- & -- & -- \\
  16 &\small{sober, beers, drink} & -- & -- & -- &  \small{youtubes, youtube, youtubers} & -- & -- & -- \\
  17 & \small{corvette, lamborghini, bentley} & -- & -- & -- &  \small{denier, congressperson, reelection} & \checkmark  & -- & \checkmark \\
  18 & \small{numbers, numerals, staggering} & -- & -- & -- &  {\small fbi, fbis, disband} & -- & -- & -- \\
  19 & \small{dangs, bagot, shrugs} & -- & -- & -- &  {\small border, borders, crossings} & -- & -- & \checkmark \\
  20 & \small{podcasting, podcasts, podcaster} & -- & -- & -- &  {\small science, scientific, scientists} & -- & -- & \checkmark \\
  \bottomrule
  \end{tabular}
  \caption{Comparison of the top 20 topics on YouTube and Rumble. The presence of a checkmark signifies that the topic appears in the top 20 topics of baseline political podcasts. }
  \label{tab:comparison_top_topics}
\end{table*}

\descr{Ethics guidelines.}
Our project, which exclusively uses publicly accessible data and does not involve human subjects, is not classified as human subjects research according to the guidelines of our institution's Institutional Review Board (IRB). 
We adhere to established ethical standards in social media research and the application of shared measurement data. 
Additionally, we only use third-party models with publicly available licences. 
We do not anonymize people if they are public figures (i.e., podcast channel owners on YouTube or Rumble).
All quotes are provided verbatim from the automated transcriptions and are attributed to the podcast and/or the person they came from.

\section{Is there a political bias in the videos of podcast channels on Rumble?}
\label{sec:rq1}

To explore political bias in Rumble podcasts, we perform a quantitative analysis using speech-to-text transcriptions.
Initially, we examine political orientations by comparing the popular topics on \dsrumble and \dsyoutube with those in \dspolitical.
We aim to determine if the discussions align with those typically found on channels known for their political activism or ideological bias, establishing a foundational understanding of the political characteristics inherent in the analyzed content.

Subsequently, we examine semantic similarities across topics using transformer-based sentence embeddings, which allow us to facilitate a deeper inference of potential political alignments or biases present within the discourse.
Our analysis extends to channel-based political stances, where we evaluate the political leanings of the podcast videos from channels on \dsrumble and \dsyoutube.
This broader perspective helps us understand the diversity of political views on these platforms and whether there is a tendency towards certain political ideologies.

\descr{Topic model.}
We use BERTopic~\cite{grootendorst2022bertopic}, a transformer-based topic modeling technique, in conjunction with MPNetv2 embeddings to extract meaningful topics used by \dsrumble and \dsyoutube, and \dspolitical.
We use this combination because of its ability to discern semantic similarities and differences among documents~\cite{hanleyHappenstanceUtilizingSemantic2023,yang2023analyzing}.
In line with prior research~\cite{hanleyHappenstanceUtilizingSemantic2023}, we split transcripts into sentences and extract their embeddings using MPNEt-base-v2 model. %
Our manual inspection of transcripts finds that 2\% of all podcast videos in our dataset are missing punctuation. For these specific transcriptions, we split the speech-to-text outputs into sentences using a model~\cite{guhrfullstop}, which achieves an F1 score of 0.94 for predicting sentence endings in English text.
Given that our analysis relies on sentences extracted from speech-to-text transcripts, which often include common spoken utterances (e.g., ``Hello everyone!''), we apply postprocessing to focus on more substantive topics, as detailed in Appendix A.

\descr{Examining the political alignment of topics.}
To identify political bias in \dsrumble and \dsyoutube, we initially assess the extent to which the topics they focus on align with those in \dspolitical.
To achieve this, we compare the most popular topics of \dsrumble and \dsyoutube with those of \dspolitical.
Table~\ref{tab:comparison_top_topics} presents the top 20 topics of \dsrumble and \dsyoutube.
A checkmark indicates if a topic also appears among the top 20 topics in a political podcast sample, where a topic can appear in more than one political leaning.

Among the popular topics of \dsrumble, 70\% align with \dsright, 50\% with \dsleft, and 40\% with \dscenter.
We find that \dsrumble focuses primarily on topics heavily discussed in politics or those that can be attributed to political discussions, with a few exceptions (topics \#6, \#12, \#15, \#16, and \#20), which are related to social media, books, science, and mundane conversations.

In contrast to \dsrumble, our analysis shows that \dsyoutube has less alignment with political spectrums, aligning 25\%, 20\%, and 30\% with \dsright, \dsleft, and \dscenter, respectively.
This indicates a reduced focus on political subjects overall.
Instead, \dsyoutube tends to feature content centered around more apolitical life interests, e.g., sports (topics \#2 and \#9), sport cars (\#17), or music (\#15).
We also note that, while \dsyoutube's most popular topics are generally more mainstream than political, the presence of  topics related to the Russian invasion of Ukraine (\#6) and
Roe v Wade overturn decision (\#12), masks (\#14), and vaccines (\#10)  suggest that popular podcast channels of YouTube can also facilitate discussions around political and social issues.

\descr{Semantic alignments with political podcasts.}
We explore the semantic similarities between \dsrumble and \dsyoutube, as well as their relationship to \dspolitical. 
Using MPNet sentence embeddings, our analysis involves performing a layered examination of semantic similarities across varying levels of topic prevalence.
Our rationale for this approach is based on an observation made during our earlier analysis, where we noted that a holistic comparison results on high similarity scores, possibly due to occurences of mundane conversations common in many podcast videos.
So, we perform our analysis beginning with the top 20 topics and expanding exponentially across five tiers, from 20 to 320 topics. 
This approach allows us to examine the semantic alignment across different tiers of topic frequency, covering nearly 20\% of the sentences in \dsrumble and \dsyoutube after postprocessing (See Figure~\ref{fig:cdf_topic_ranks} in the Appendix). 
Nonetheless, we also present the semantic similarities that cover all topics.

To determine semantic similarities between the datasets, we first calculate the centroids of the top N topics for each dataset. 
The semantic similarity is then assessed using the cosine similarity between these centroids for the top N topics of each dataset.
This method provides a nuanced view of the semantic connections between \dsrumble and \dsyoutube in comparison to \dspolitical, across multiple strata of topic concentration.

\begin{figure}[th]
  \centering

  \begin{subfigure}[b]{0.9\linewidth}
    \includegraphics[width=\linewidth]{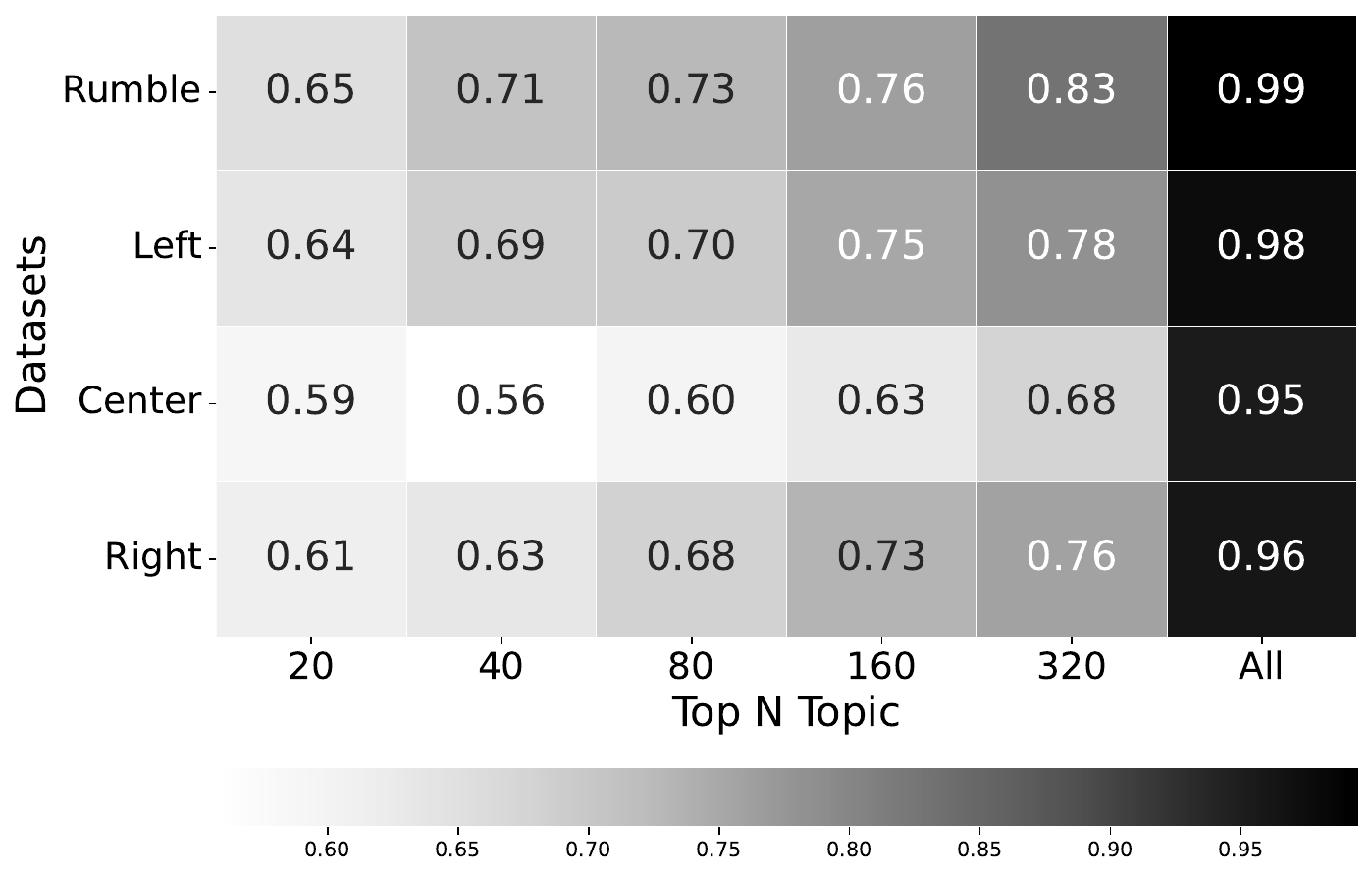}
    \caption{YouTube}
    \label{fig:heatmap_YouTube_refined_semantic}
\end{subfigure}
\hfill
  \begin{subfigure}[b]{0.9\linewidth}
      \includegraphics[width=\linewidth]{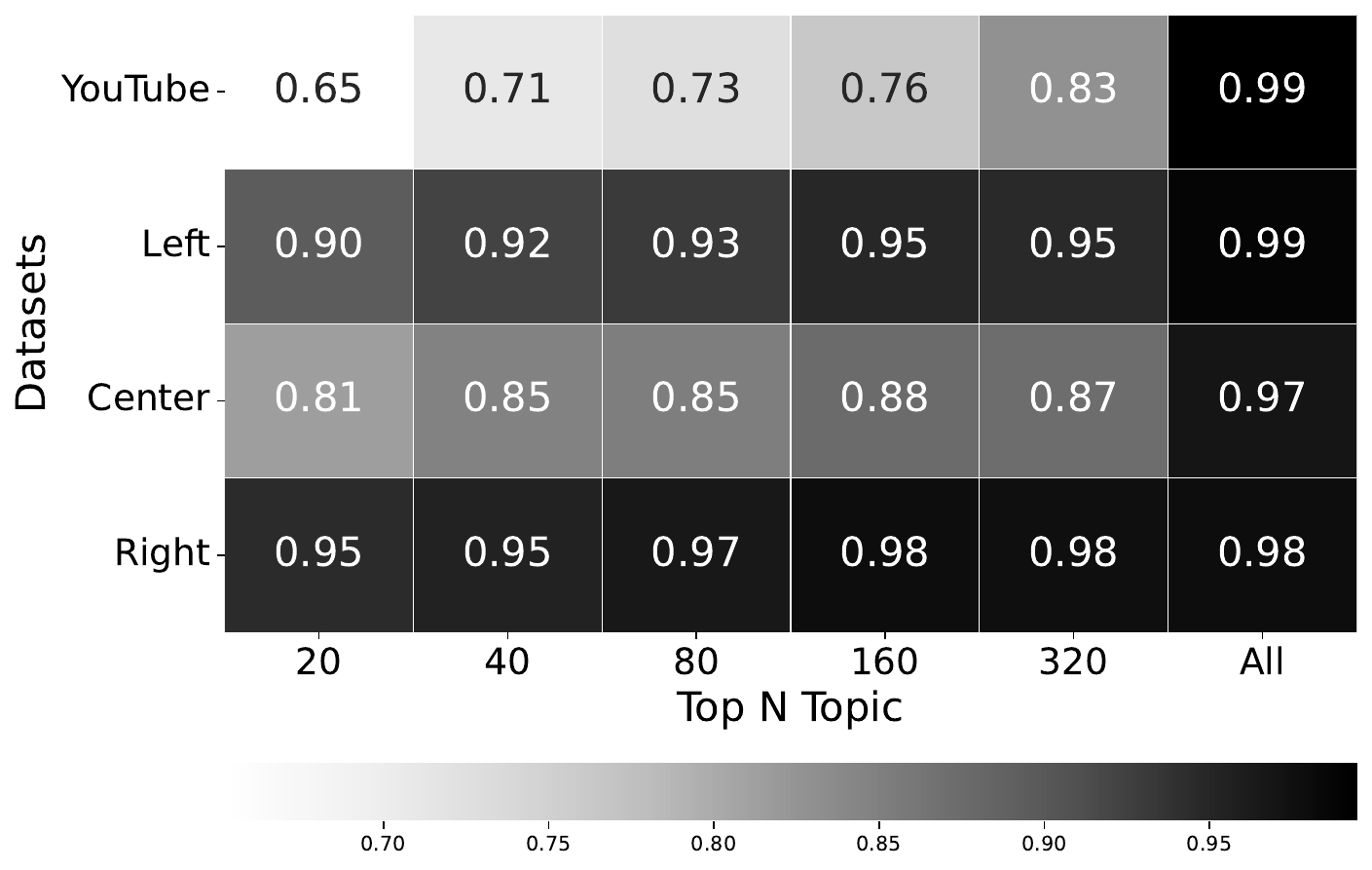}
      \caption{Rumble}
      \label{fig:heatmap_rumble_refined_semantic}
  \end{subfigure}
  \caption{Heatmaps illustrating the cosine similarities among the top N topic centroids: (a)~YouTube versus Rumble, and across left-wing, center, and right-wing podcasts; (b)~Rumble versus YouTube, and across left-wing, center, and right-wing podcasts. Darker shades denote greater semantic similarity.}
  \label{fig:heatmaps}
\end{figure}

As seen in Figure~\ref{fig:heatmap_rumble_refined_semantic}, \dsrumble exhibits similarity scores of $\geq 0.95$ with \dsright across all ranks.
In comparison, the similarity scores are $\geq 0.90$ with \dsleft, $\geq 0.81$ with \dscenter, and $\geq 0.65$ with \dsyoutube.
These results indicate a high semantic alignment within Rumble's podcast videos.
However, this high alignment causes \dsrumble's semantic relationships with \dspolitical to appear more closely aligned than they might actually be.
To address this, we normalized Rumble's semantic similarities with the \dspolitical datasets.
This adjustment helps eliminate the influence of non-political content in these similarities, providing a more detailed understanding of \dsrumble's semantic similarity with political content.
As a result, we find that \dsrumble has similarity scores of $\geq 0.75$ with \dsright across all ranks, compared to $\geq 0.47$ with \dsleft.
Further details are provided in Figure~\ref{fig:rumble_semantic_normalized} in the Appendix.

When we look at Figure~\ref{fig:heatmap_YouTube_refined_semantic}, we see considerably lower semantic similarities between \dsyoutube and \dspolitical.
Furthermore, we find that \dsrumble shows less similarity with \dsyoutube compared to \dsrumble's semantic similarities with \dspolitical.
Although these similarity scores increase with the topic size, it is evident that \dsrumble shows more pronounced semantic similarities with \dspolitical, particularly with \dsright, in contrast to \dsyoutube.

\begin{figure}[t]
  \centering
  \includegraphics[width=0.98\linewidth]{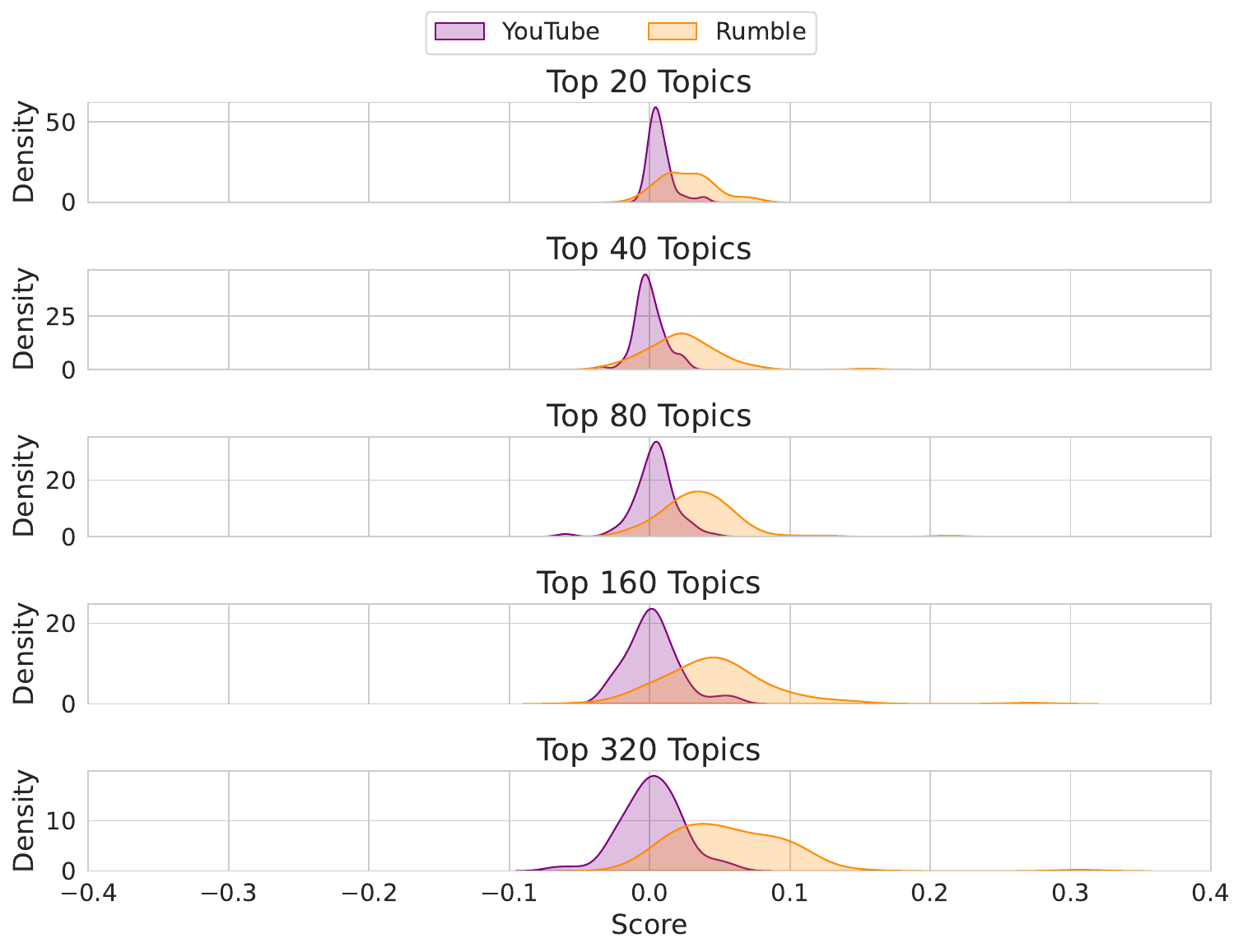}
  \caption{Density plots of political alignment scores for Rumble and YouTube channels. Scores represent ideological orientation and range from -1 to 1, where negative values denote a left-leaning bias and positive values suggest a right-leaning inclination.}
\label{fig:density_similarity}
\end{figure}

\begin{figure}[t]
  \centering
  \includegraphics[width=0.98\linewidth]{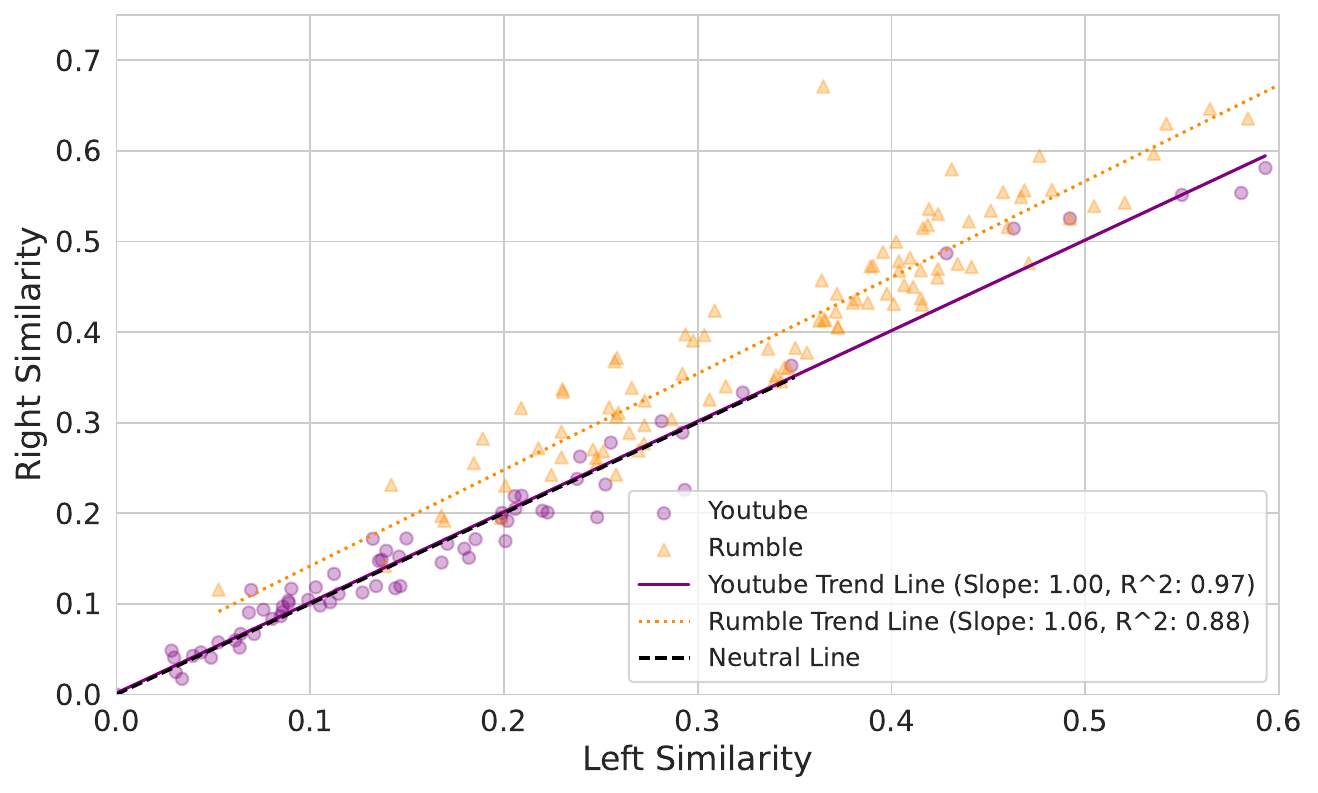}
  \caption{Scatter plot showing right-wing and left-wing similarity distributions for the top 320 topics in podcast videos of popular YouTube and Rumble podcast channels, with R-squared and slope values from linear regression.}
\label{fig:regression}
\end{figure}

\descr{Channel-based ideological alignments.}
Last, we compare the channel-level similarities between \dsrumble and \dsyoutube with \dsleft and \dsright.
To measure their similarities, we calculate the percentage of intersecting topics.
Specifically, for the podcast videos of each channel on \dsrumble and \dsyoutube, we identify the top 320 topics and evaluate their intersection with the top \( N \) topics of \dsleft and \dsright, where \( N \) increments exponentially from 20 to 320.
This method allows us to assess the breadth of topics covered by the podcast videos of each channel on \dsrumble and \dsyoutube and how they intersect with the political spectrum at various levels. 
By exponentially increasing $N$ for the \dsleft and \dsright, we can measure how their content aligns or diverges from the broader topic set of \dsrumble and \dsyoutube.

To quantify this similarity, we compute the difference in intersection percentages with \dsleft and \dsright topics:

\[
SimScore_{T_i} = \frac{| C \cap R_{T_i} | - | C \cap L_{T_i} |}{| C |} 
\]

where \( C \) represents the set of topics of a given channel, while \( R_{T_i} \) and \( L_{T_i} \) correspond to the top \( T_i \) topics from \dsleft and \dsright, respectively. 
For our purposes, \( C\) is the set of the top 320 topics, and $T_i = \{20, 40, 80, 160, 320\}$.

Figure~\ref{fig:density_similarity} plots density distributions of political similarity scores for \dsrumble and \dsyoutube.
Complementing this, Figure~\ref{fig:regression} displays the distribution of left-wing and right-wing similarity scores for the top 320 topics in \dsrumble and \dsyoutube channels.
This scatter plot also includes R-squared and slope values derived from linear regression analysis, providing further details into the patterns observed in Figure~\ref{fig:density_similarity}. 
It is evident that \dsyoutube predominantly clusters around the neutral score (0) across all top N topics, whereas \dsrumble exhibits a distribution skewed towards the right-wing, indicated by predominantly positive (right-wing leaning) maximum densities.
This is another indication of Rumble podcasts' overall right-wing political leaning.

\descr{Takeaways.} 
Rumble's popular podcasts lean predominantly towards political topics. 
Our analysis reveals that this leaning is evident not only in the platform's overall content but also in channel-wise leanings. 
Their topical focus aligns the most with right-wing podcasts, where we also find Rumble has more than 95\% semantic alignment with right-wing podcast content.
Moreover, there is a clear inclination towards right-wing content at the channel level. 
This contrasts with YouTube, where podcasts have a broader focus, covering a wide array of mainstream topics and interests beyond the political sphere. 
Our results are further supported when we compare the word usages between \dsyoutube and \dsrumble (detailed in Appendix B), where we find that \dsrumble aligns with general right-wing narratives on topics related to abortion, BLM protests, and the January 6 Capitol attack.

\begin{table*}
  \centering
  \scriptsize
    \begin{tabular}{rlr|rlr}
    \toprule
    \multicolumn{3}{c|}{YouTube} & \multicolumn{3}{c}{Rumble} \\
    \midrule
  
          no. &	Top 3 Topic Words &  Coefficient & no. &	Top 3 Topic Words &  Coefficient  \\
  \midrule
  1&\small{bear, grizzly, bears}             &        1,722,484 & 1&\small{canceled, cancel, canceling} &       339,579   \\
  2&\small{shots, shot, twoshot}           &        811,020  & 2&\small{omicron, omnicron, variant}           &        171,452   \\
  3&\small{monkeys, chimps, apes}         &        736,789  & 3&\small{machines, machine, machinery} &        130,210   \\
  4&\small{understand, craig, dirk}    &        496,204 & 4&\small{ivermectin, antiparasitic, deworm} &        99,353 \\
  5&\small{nasa, astronaut, moon}          &        481,355  & 5&\small{pharma, fda, pharmaceutical} &        89,163 \\
  6&\small{pizza,pizzas,hut} &        470,895  & 6&\small{check, doozy, peter} &        86,759 \\
  7&\small{jew, jewish, jews}                &        446,220  & 7&\small{sober, beers, drink} &        86,384 \\
  8&\small{jail, juvenile, prison} &        414,254 & 8&\small{study, studies, metaanalysis} &       84,177\\
  9&\small{insulin, glucose, diabetes}                  &        413,936& 9&\small{platforms, platform, cdp} &        77,112  \\
  10&\small{fish, fishing, trout}       &        413,030  & 10&\small{joes, joe, sleepy} &        76,842 \\
    
  \bottomrule
  \end{tabular}
  \caption{Top topics on YouTube and Rumble that show a significant ($p \leq 0.001$) positive correlation with viewership.}
  \label{tab:table_view_topic}
  \end{table*}

\section{What content drives the number of views on popular Rumble and YouTube podcast channels?}
\label{sec:rq2}

We use linear regression to analyze the correlation between the topics and their number of views for \dsrumble and \dsyoutube.
This method allows us to capture the relationship between the specific topical focuses and user engagement, particularly emphasizing a small number of videos that have a high number of views in certain topic categories.
Consistent with previous experiments, our analysis is concentrated on the top 320 topics from both \dsrumble and \dsyoutube to focus on the most engaging discussions.
Additionally, we select a sample of 100 sentences for manual inspection to better understand the content within these topics.

Table~\ref{tab:table_view_topic} presents the top 10 topics that have a statistically significant correlation ($p \leq 0.001$) with the podcast video view counts, along with their respective coefficients. 
In our results, we removed three topics after identifying them as commercials.
We find that the discussions centered around ``cancellation'' exhibit the highest correlation with \dsrumble views.
Next, we take a closer look into the top 20 channels that have the highest number of sentences related to cancellation.
We examine the number of podcast videos they are associated with and the average views these podcast videos receive. 
The result is detailed in Table~\ref{tab:cancelled_info} in the Appendix.

We discover that Andrew Tate's channel has the highest average number of views within the ``cancellation'' topic.
Interestingly, when we remove this channel from our analysis, the correlation is no longer statistically significant, which indicates that discussions about the cancellation of a prominent right-wing figure have a positive and statistically significant impact on Rumble's view counts.
Moreover, we find a moderate positive correlation (Spearman's, $\rho = 0.5$, $p \leq 0.001$) in the rankings of other topics after excluding the ``TateSpeech by Andrew Tate'' channel from our analysis.
This finding suggests that Andrew Tate's presence alone has a significant impact on the podcast views of Rumble.
Besides the conversations focused on cancellation, we also find that discussions related to COVID-19 Omicron variant (\#2), Ivermectin (\#4), and the FDA (\#5) are among the top 10 topics correlated with views. 
These topics have been central to the discourse during the Covid-19 pandemic~\cite{CollinsZadrozny2021}.
An example sentence from  the ``Viva Frei:''

\begin{quote}

  \textit{``And the, the FDA has designed their sets of rules, the way the Biden administration designed all their mandates.''}
  \end{quote}

We mainly encounter mundane discussions on topics \#3, \#6, and \#7.
In the topic related to academic studies (\#8), we encounter discussions on various academic studies.

We find the discussions on platform topic (\#9) is mostly centered around discussions on social media platforms.
We encounter many sentences regarding alternative platforms:
An example sentence from  ``Dinesh D'Souza:''

\begin{quote}

  \textit{``If we're thrown off digital platforms, we have to create our own platforms.''}
  \end{quote}

Moreover, we encounter sentences using the derogatory nickname ``Sleepy Joe'' for Joe Biden on topic \#10.
An example sentence from ``Rob Carson Show podcast:''

\begin{quote}

  \textit{``And all of a sudden, Sleepy Joe, he's looking strong and dependable without doing anything particularly inspiring.''}
  \end{quote}

In our analysis of the top 10 topics of \dsyoutube that are significantly correlated with their view count, we find fewer meaningful topics.
Initially, the six topics with the highest correlation are related to the names, e.g., ``Sam'' and ``Matt.'' where further examination of the channels with the most sentences on these topics reveals that this is attributed to podcast hosts, e.g.,  Sam Seder from ``The Majority Report w/ Sam Seder'' and Sam Riegel from ``Critical Role.''
This might suggest that the primary motivation for YouTube podcast viewers is listening to the discussions led by specific hosts rather than engaging with particular topics. 
To provide a more nuanced analysis, we subsequently removed these host-specific topics from our results.

Consistent with our previous findings, the topics on YouTube that are most strongly correlated with view counts are predominantly apolitical (e.g., animals (\#1 and \#3), space (\#5), pizza (\#6), and fishing (\#10)).
We also encounter topics related to prison (\#8) and diabetes (\#9).
An example sentence from ``Fresh and Fit:''

\begin{quote}

  \textit{``So really the thing with me was when I went, I went to prison for two years.''}
  \end{quote}

However, we also identify a topic concerning Jews, which might be politically motivated (\#7). 
Upon manually inspecting the sentences within this topic, we find that they frequently focus on Jewish identity.
We find that H3 Podcast primarily drives this discussion with 204 sentences from 45 podcasts:
\begin{quote}

  \textit{``I feel like people make that characterization against me because I'm Jewish.''}
  \end{quote}

Despite encountering antisemitic statements, our manual review indicates that these are satirical comments made by the Jewish hosts of the channel:
\begin{quote}

  \textit{``No, I hate Jews.''}
  
  \end{quote}

Ironically, this channel was previously suspended from YouTube due to a joke about the Holocaust~\cite{Lapin}.
H3 Podcast was also banned from YouTube for 7 days for making a joke about bombing the NRA after the elementary school shooting in Uvalde, Texas~\cite{H3bomb}.
H3 Podcast has also been sued for even seemingly innocent issues, e.g., providing helpful services~\cite{weinstein_2024} to make sure people are not falsely being confused with Harvey Weinstein~\cite{Fisher2023}.

\begin{table*}[th!]
  \centering
  \footnotesize
  \begin{tabular}{rlccc|rlccc}
  \toprule
  \multicolumn{5}{c|}{YouTube} & \multicolumn{5}{c}{Rumble} \\
  \midrule
  \textbf{no.} & \textbf{Label | (\% Channels)} & \textbf{Left} & \textbf{Center} & \textbf{Right} & \textbf{no.} & \textbf{Label | (\% Channels)} & \textbf{Left} & \textbf{Center} & \textbf{Right} \\
  \midrule
  1 & \scriptsize{Captioned images -- (46)} & -- & -- & \checkmark                       & 1 &  \scriptsize{Joe Biden -- (34)}  & -- & -- & -- \\
  
  2 & \scriptsize{Guests (Video conference) -- (24)}   & \checkmark & -- & \checkmark                               & 2 &  \scriptsize{Jen Psaki -- (31)}  & \checkmark & -- & \checkmark \\
  
  3 & \scriptsize{Smart Phones -- (21)} & -- & -- & --                              & 3 &  \scriptsize{Covid-19 News -- (31)}                    & -- & -- & \checkmark \\
  
  4 & \scriptsize{Cartoons -- (19)}            & -- & -- & --                               & 4 &  \scriptsize{Hillary Clinton -- (31)}                      & -- & -- & -- \\
  
  5 & \scriptsize{Nostalgic Photos -- (17)}  & -- & -- & --                               & 5 &  \scriptsize{Ron Desantis -- (30)}                 & -- & -- & \checkmark \\
  
  6 & \scriptsize{Basketball Court -- (17)} & -- & -- & --                      & 6 &  \scriptsize{Kamala Harris -- (29)} & -- & -- & \checkmark \\
  
  7 & \scriptsize{Google Image Queries -- (16)}    & -- & -- & --                               & 7 &  \scriptsize{Guests (Video conference) -- (28)}              & \checkmark & -- & \checkmark \\
  
  8 & \scriptsize{Typing (keyboard) -- (16)} & -- & -- & --                                & 8 &  \scriptsize{Canadian Politics -- (27)}                     & -- & -- & \checkmark \\
  
  9 & \scriptsize{Space -- (16)}                & -- & -- & --                                & 9 &  \scriptsize{Captioned images -- (24)}                 & -- & -- & \checkmark \\
  
  10 & \scriptsize{Podcast Studio -- (14)} & -- & -- & --                               & 10 &  \scriptsize{Tucker Carlson -- (23)}               & -- & -- & -- \\
  
  11 & \scriptsize{Joe Rogan -- (14)}         & -- & -- & --                                & 11 &  \scriptsize{Joe Biden (w/ mask) -- (23)}                         & -- & -- & -- \\
  
  12 & \scriptsize{Money -- (14)} &-- &-- & --                                 & 12 &  \scriptsize{Rand Paul -- (22)}       & -- & -- & -- \\
  
  13 & \scriptsize{Typing (smart phone) --  (14)} & -- & -- & --                               & 13 &  \scriptsize{Anthony Fauci -- (21)}                   & -- & -- & \checkmark \\
  
  14 & \scriptsize{Science -- (13)} & -- & -- & --                                   & 14 &  \scriptsize{Whoopi Goldberg -- (21)}                    & -- & -- & -- \\
  
  15 & \scriptsize{Instagram -- (13)} & -- & -- & --                                & 15 &  \scriptsize{Karine Jean-Pierre
  -- (20)}                     & -- & -- & --\\
  
  16 & \scriptsize{Fire Images -- (13)} & -- & -- & --                               & 16 &  \scriptsize{Joe Biden (News) -- (19)}                  & -- & -- & -- \\
  
  17 & \scriptsize{Kardashians -- (13)} & -- & -- & --                                & 17 &  \scriptsize{Gavin Newsom -- (19)}           & -- & --& -- \\
  
  18 & \scriptsize{Animals -- (13)} & -- &-- & --                                  & 18 &  \scriptsize{Press conference -- (19)}                     & -- & -- & -- \\
  
  19 & \scriptsize{Photographers -- (13)} & -- & -- & --                                 & 19 &  \scriptsize{Joe Rogan-- (19)}                 & -- & -- & -- \\
  
  20 & \scriptsize{Clocks -- (13)} & -- & -- & --                                 & 20 &  \scriptsize{Bill gates -- (19)}                       & -- & -- & -- \\
  \bottomrule
  \end{tabular}
  \caption{Comparison of the Top 20 visual clusters detected through image clustering (manually labelled) on YouTube and Rumble. The presence of a checkmark signifies that the topic appears in the top 20 visual themes of left-wing, center, or right-wing podcasts.}
  \label{tab:comparison_top_images}
\end{table*}

\descr{Takeaways.}
Our analysis reveals distinct patterns in user engagement across Rumble and YouTube. 
On Rumble, there is a pronounced trend where topics related to controversial COVID-19 content and platform cancellation of right-wing figures are key drivers of podcast video views. 
This suggests a strong resonance between Rumble's content offerings and the interests of its audience in right-leaning political themes.
In contrast, YouTube present a markedly different scenario. 
Here, it is the more universal, mainstream topics that align with a broader spectrum of interests that predominantly attract views. 
This indicates YouTube's diverse appeal, suggesting that its content moderation strategy caters to a wider, less politically-focused audience.

\section{What are the most widely used visual elements? Do they share commonalities with politically motivated podcasts?}
\label{sec:rq3}

Similar to Rumble, the literature on the usage of visual elements in podcasts is also relatively scant (see Introduction).
Recognizing this gap, our analysis focuses on the visual topics covered in podcasts videos.
By examining these visual topics, we aim to have a foundational understanding on how podcasts on Rumble use visual strategies beyond mere auditory content.
Based on our previous results, we hypothesize that podcasts on Rumble also use politically motivated visual elements that align with those found in right-wing podcasts.
To investigate this, we first extract representative video frames from the podcast videos.
Subsequently, we apply a clustering technique to identify and analyze the visual clusters that are most frequently used in \dsrumble and \dsyoutube channels.

\descr{Extracting representative video frames.}
To effectively analyze the visual clusters, our first step is to extract representative video frames. 
This approach helps us avoid clusters of sequential and almost identical images from the same video.
We begin by extracting frames from each podcast video at a rate of one frame per second.
Adopting a technique used in previous research~\cite{zannettou2018origins}, we first apply perceptual hashing (pHash) to each sampled frame. 
This method extracts representative feature vectors from the images, capturing their visual characteristics. 
We then measure the similarity between frames by calculating the Hamming distance and set a threshold to identify frames with meaningful visual differences.
To establish this threshold, we tested 20 sample videos from both \dsrumble and \dsyoutube.
Starting with the second frame, we eliminate frames that fell below a varying threshold $\theta$ compared to any of the previous video frames, ranging from $\theta = 5$ to $\theta = 50$ in increments of 5. 
This evaluation is conducted by three authors of this paper who individually analyze the extracted frames for each sampled video at each $\theta$ level, focusing on two metrics:
1)~minimizing the number of duplicate images, and
2)~maximizing the number of visually distinct images.
In the end, the authors reached a unanimous agreement (Fleiss’ Kappa 1.0) on setting the threshold at $\theta = 20$.

\descr{Finding clusters of widely used visual elements.}
After extracting representative images from podcast videos, we apply hierarchical clustering to embeddings generated by OpenAI's CLIP~\cite{radford2021learning} (see details in Appendix C).
To determine the most commonly used visual clusters across various channels, we start by identifying the clusters that appear in the highest number of channels for each dataset.
Starting from the highest ranked clusters for each dataset, three authors of this paper examine 20 randomly sampled images (or the entire set if a visual cluster comprised $\leq 20$ images) and labeled clusters based on their domain knowledge.
This process is repeated until we have a definitive list of the top 20 visual clusters for each dataset, where we do not include clusters that are primarily composed of frames without meaningful visual content (e.g., black screens or solid colors, including those showing only a channel logo).

\descr{Top visual clusters of Rumble and YouTube.} 
Table~\ref{tab:comparison_top_images} displays the most frequently used visual clusters across \dsrumble and \dsyoutube, and their alignments with those in \dspolitical.
Figure~\ref{fig:top10_cluster} shows top-10 clusters for each platform.
For \dsrumble, we observe that the most prevalent visual clusters align with our earlier findings, focusing predominantly on political figures.
Notably, while the majority of politicians are associated with the left-wing (e.g., Joe Biden, Kamala Harris, and Hillary Clinton), we also see politicians and political commentators that are recognized for their right-wing perspectives (i.e., Tucker Carlson, Ron DeSantis, and Rand Paul).
We also observe that the majority of the visual elements on Canadian politics topic are related to Justin Trudeau.
We observe many anti-vaccine related news on Covid-19 News topic.
Additionally, we encounter a visual topic related to Anthony Fauci, the former Chief Medical Advisor to the President during the COVID-19 pandemic, who has been a target of criticism from right-wing figures, including former President Trump himself~\cite{CollinsLiptak2020}.
Interestingly, Bill Gates also appeared among top 20 visual clusters of Rumble, who has been at the center of COVID-19 related conspiracy theories deployed by the right-wing~\cite{mcneil2022understanding,Wakabayashi2020}.
Comparing these findings with the top 20 visual clusters from \dsleft, \dsright, and \dscenter, we find alignments of 10\%, 40\%, and 0\% respectively. 
This suggests that Rumble's podcasts exhibit meaningfully more visual commonalities with right-wing podcasts.

Our results from \dsyoutube's most widely used visual clusters also aligns with our previous findings, as these visuals consist of mostly apolitical and more mainstream themes (e.g., cartoons, basketball court, and Kardashians).
When comparing these results to the top 20 most widely used visual clusters in \dspolitical, we find 5\% alignment with \dsleft, 10\% with \dsright, and no alignment (0\%) with \dscenter.

\descr{Takeaways.}
Rumble podcasts' visual content is primarily political, with popular visual clusters aligning closely with right-wing podcasts.
We observe that these clusters predominantly feature political figures. 
While these clusters largely showcase left-wing politicians, the political commentators within them are typically associated with right-wing viewpoints.
One possible explanation for this could be the dominance of the Democratic Party in the US government during the majority of our dataset's timeline. 
This may suggest that Rumble's podcasts use visuals of these politicians while critiquing them, stimulating their viewers beyond merely using audio. 
On YouTube, we consistently find a dominance of apolitical visual clusters, aligning with our prior observations. 
This contrast further underscores Rumble's non-neutral political stance.

\section{Discussion \& Conclusion}

In this paper, we analyze the audio-visual content of popular Rumble and YouTube podcast channels, focusing on their political leanings.
Our analysis of over 13K podcast videos demonstrates a right-wing bias in Rumble's content, which sharply constrasts with YouTube's more apolitical content. 
This dichotomy highlights the role of platforms in either reinforcing or challenging existing political narratives. 
Our findings suggest that Rumble's video podcast content is predominantly right-wing content, thereby creating a distinct echo chamber effect~\cite{efstratiou2022non}. 
This phenomenon is critical to understand, as it potentially exacerbates societal polarization in yet underexplored area, e.g., podcasts.

Our findings also emphasize the need to consider both audio and visual elements in media studies.
While textual content has been extensively analyzed in social media research, our findings reveal that audio-visual content in podcasts can reinforce polarized political beliefs in social media platforms.
The use of specific visual elements, which align with the themes commonly associated with right-wing misinformation, further intensifies the impact of these podcasts on political engagement.
Furthermore, our analysis reveals that controversial content related to COVID-19, and the deplatforming of right-wing figures are significant drivers of podcast video views on Rumble, indicating alternative controversial discussions/figures are the drivers of podcast views on Rumble.

\subsection{Implications \& Future Work}

Future research should expand to other alternative video platforms, e.g., BitChute, to compare their video podcast content with Rumble and YouTube, and to understand the broader political video podcast content across various platforms. 
Studies could examine how the political bias in audio-visual content in these platforms evolves over time, and affect each other, particularly in response to major political events or changes in platform policies.
Moreover, similar to studies that analyze deplatformed communities on social media~\cite{horta2021platform, jhaver2021evaluating, balci2024exploring, ali2021understanding, balci2023beyond, patel2024idrama}, future work could analyze the effect of deplatforming on the podcast hosts themselves.
Exploring the algorithms used by these alternative video sharing platforms to recommend content is another crucial area for future research. 
Understanding the mechanics behind podcast video content recommendation on alternative video sharing platforms like Rumble could reveal insights into how and why certain political content is amplified.
Finally, integrating audio-visual content analysis of podcast videos with other media forms like text-based social media, news articles, and TV broadcasts, could offer a more comprehensive picture of the podcast video ecosystem and its influence on political polarization.

\subsection{Limitations}

This work is subject to certain limitations. 
First, the data collection was not conducted live, which means some content may have been missed.
Furthermore, as we rely on content creators' labeling to create our initial set of podcast videos, our methodology might miss some podcast videos that are not labeled by the creators.
Moreover, our reliance on tools like faster-whisper, BERTopic, and CLIP, could introduce errors due to their inherent limitations, e.g., Whisper is known for hallucinating content~\cite{mittal2024towards,koenecke2024careless} and BERTopic can generate higher number of outliers than expected~\cite{egger2022topic}.
These factors should be considered when interpreting our findings.
Additionally, our analysis faces other limitations. 
For instance, our labeling of the representative visual elements in Rumble and YouTube podcasts was guided by our domain knowledge, yet some channel owners might challenge our categorizations.
Another limitation of our study involves assessing how the content of Rumble and YouTube podcasts aligns with political orientations, without analyzing the sentiment of this content. 
While this methodology was in line with our research objectives, it is important to recognize that including sentiment analysis might have offered additional insights into the emotional tone and impact of the podcast content.

\section{Acknowledgments}

This material is based upon work supported by the National Science Foundation under Grant No. IIS-2046590.


\begin{thebibliography}{85}
\providecommand{\natexlab}[1]{#1}

\bibitem[{Ali et~al.(2021)Ali, Saeed, Aldreabi, Blackburn, De~Cristofaro,
  Zannettou, and Stringhini}]{ali2021understanding}
Ali, S.; Saeed, M.~H.; Aldreabi, E.; Blackburn, J.; De~Cristofaro, E.;
  Zannettou, S.; and Stringhini, G. 2021.
\newblock Understanding the effect of deplatforming on social networks.
\newblock In \emph{WebSci}, 187--195.

\bibitem[{Aliapoulios et~al.(2021{\natexlab{a}})Aliapoulios, Bevensee,
  Blackburn, Bradlyn, De~Cristofaro, Stringhini, and
  Zannettou}]{aliapoulios2021early}
Aliapoulios, M.; Bevensee, E.; Blackburn, J.; Bradlyn, B.; De~Cristofaro, E.;
  Stringhini, G.; and Zannettou, S. 2021{\natexlab{a}}.
\newblock An early look at the parler online social network.
\newblock \emph{arXiv:2101.03820}.

\bibitem[{Aliapoulios et~al.(2021{\natexlab{b}})Aliapoulios, Bevensee,
  Blackburn, Bradlyn, De~Cristofaro, Stringhini, and
  Zannettou}]{aliapoulios2021large}
Aliapoulios, M.; Bevensee, E.; Blackburn, J.; Bradlyn, B.; De~Cristofaro, E.;
  Stringhini, G.; and Zannettou, S. 2021{\natexlab{b}}.
\newblock A large open dataset from the Parler social network.
\newblock In \emph{ICWSM}, volume~15, 943--951.

\bibitem[{Ammar et~al.(2016)Ammar, Mulcaire, Tsvetkov, Lample, Dyer, and
  Smith}]{ammar2016massively}
Ammar, W.; Mulcaire, G.; Tsvetkov, Y.; Lample, G.; Dyer, C.; and Smith, N.~A.
  2016.
\newblock Massively multilingual word embeddings.
\newblock \emph{arXiv:1602.01925}.

\bibitem[{Angelov(2020)}]{angelov2020top2vec}
Angelov, D. 2020.
\newblock Top2vec: Distributed representations of topics.
\newblock \emph{arXiv:2008.09470}.

\bibitem[{Balci et~al.(2023)Balci, Ling, De~Cristofaro, Squire, Stringhini, and
  Blackburn}]{balci2023beyond}
Balci, U.; Ling, C.; De~Cristofaro, E.; Squire, M.; Stringhini, G.; and
  Blackburn, J. 2023.
\newblock Beyond Fish and Bicycles: Exploring the Varieties of Online Women’s
  Ideological Spaces.
\newblock In \emph{WebSci}, 43--54.

\bibitem[{Balci et~al.(2024)Balci, Patel, Balci, and
  Blackburn}]{balci2024idrama}
Balci, U.; Patel, J.; Balci, B.; and Blackburn, J. 2024.
\newblock iDRAMA-rumble-2024: A Dataset of Podcasts from Rumble Spanning 2020
  to 2022.
\newblock \emph{Workshop Proceedings of the 18th International AAAI Conference
  on Web and Social Media}.

\bibitem[{Balci, Sirivianos, and Blackburn(2024)}]{balci2024exploring}
Balci, U.; Sirivianos, M.; and Blackburn, J. 2024.
\newblock Exploring Left-Wing Extremism on the Decentralized Web: An Analysis
  of Lemmygrad. ml.
\newblock \emph{Workshop Proceedings of the 18th International AAAI Conference
  on Web and Social Media}.

\bibitem[{Boesinger et~al.(2024)Boesinger, Ribeiro, Veselovsky, and
  West}]{boesinger2023tube2vec}
Boesinger, L.; Ribeiro, M.~H.; Veselovsky, V.; and West, R. 2024.
\newblock Tube2Vec: Social and Semantic Embeddings of YouTube Channels.
\newblock In \emph{ICWSM}, volume~18, 2084--2090.

\bibitem[{Bratcher(2022)}]{bratcherDeeperDiscussionSurvey2022a}
Bratcher, T.~R. 2022.
\newblock Toward a Deeper Discussion: A Survey Analysis of Podcasts and
  Personalized Politics.
\newblock \emph{AJC}, 30(2): 188--199.

\bibitem[{Brown(2022)}]{Brown2022}
Brown, A. 2022.
\newblock Is rumble, a right-wing social media company, already the next meme
  stock?
\newblock
  \url{https://www.forbes.com/sites/abrambrown/2021/12/02/rumble-spac-ipo-social-media-conservative}.

\bibitem[{Chadha, Avila, and {Gil de
  Z{\'u}{\~n}iga}(2012)}]{chadhaListeningBuildingProfile2012}
Chadha, M.; Avila, A.; and {Gil de Z{\'u}{\~n}iga}, H. 2012.
\newblock Listening {{In}}: {{Building}} a {{Profile}} of {{Podcast Users}} and
  {{Analyzing Their Political Participation}}.
\newblock \emph{J. Inf. Technol. Politics}, 9(4): 388--401.

\bibitem[{Chen and Ferrara(2023)}]{chen2023tweets}
Chen, E.; and Ferrara, E. 2023.
\newblock Tweets in time of conflict: A public dataset tracking the twitter
  discourse on the war between Ukraine and Russia.
\newblock In \emph{ICWSM}, volume~17, 1006--1013.

\bibitem[{Cho, Park, and Choi(2023)}]{choMotivesUsingNews2023}
Cho, Y.~Y.; Park, A.; and Choi, J. 2023.
\newblock Motives for Using News Podcasts and Political Participation Intention
  in {{South Korea}}: {{The}} Mediating Effect of Political Discussion.
\newblock \emph{Media International Australia}, 187(1): 39--56.

\bibitem[{Clifton et~al.(2020)Clifton, Reddy, Yu, Pappu, Rezapour, Bonab,
  Eskevich, Jones, Karlgren, Carterette, and Jones}]{clifton100000Podcasts2020}
Clifton, A.; Reddy, S.; Yu, Y.; Pappu, A.; Rezapour, R.; Bonab, H.; Eskevich,
  M.; Jones, G.; Karlgren, J.; Carterette, B.; and Jones, R. 2020.
\newblock 100,000 {{Podcasts}}: {{A Spoken English Document Corpus}}.
\newblock In Scott, D.; Bel, N.; and Zong, C., eds., \emph{COLING}, 5903--5917.

\bibitem[{Collins and Zadrozny(2021)}]{CollinsZadrozny2021}
Collins, B.; and Zadrozny, B. 2021.
\newblock Clamoring for ivermectin, some turn to a pro-trump telemedicine
  website.
\newblock
  \url{https://www.nbcnews.com/tech/tech-news/ivermectin-demand-drives-trump-telemedicine-website-rcna1791}.

\bibitem[{Collins and Liptak(2020)}]{CollinsLiptak2020}
Collins, K.; and Liptak, K. 2020.
\newblock Trump trashes Fauci and makes baseless coronavirus claims in campaign
  call | CNN politics.
\newblock
  \url{https://www.cnn.com/2020/10/19/politics/donald-trump-anthony-fauci-coronavirus/index.html}.

\bibitem[{Danilak(2021)}]{langdetect}
Danilak, M.~M. 2021.
\newblock {Language detection library ported from Google's language-detection.}
\newblock \url{https://pypi.org/project/langdetect/}.

\bibitem[{Dinkov et~al.(2019)Dinkov, Ali, Koychev, and
  Nakov}]{dinkovPredictingLeadingPolitical2019}
Dinkov, Y.; Ali, A.; Koychev, I.; and Nakov, P. 2019.
\newblock Predicting the {{Leading Political Ideology}} of {{YouTube Channels
  Using Acoustic}}, {{Textual}}, and {{Metadata Information}}.
\newblock arxiv:1910.08948.

\bibitem[{Efstratiou et~al.(2022)Efstratiou, Blackburn, Caulfield, Stringhini,
  Zannettou, and De~Cristofaro}]{efstratiou2022non}
Efstratiou, A.; Blackburn, J.; Caulfield, T.; Stringhini, G.; Zannettou, S.;
  and De~Cristofaro, E. 2022.
\newblock Non-Polar Opposites: Analyzing the Relationship Between Echo Chambers
  and Hostile Intergroup Interactions on Reddit.
\newblock \emph{arXiv:2211.14388}.

\bibitem[{Egger and Yu(2022)}]{egger2022topic}
Egger, R.; and Yu, J. 2022.
\newblock A Topic Modeling Comparison Between LDA, NMF, Top2Vec, and BERTopic
  to Demystify Twitter Posts.
\newblock \emph{Frontiers in Sociology}, 7.

\bibitem[{Escandon(2024)}]{Escandon2024b}
Escandon, R. 2024.
\newblock “video podcasting” growing in popularity.
\newblock
  \url{https://www.forbes.com/sites/rosaescandon/2024/04/29/video-podcasting-growing-in-popularity}.

\bibitem[{Euritt(2019)}]{eurittPublicCirculationNPR2019}
Euritt, A. 2019.
\newblock Public Circulation in the {{NPR Politics Podcast}}.
\newblock \emph{Popular Communication}, 17(4): 348--359.

\bibitem[{Fandom(2023)}]{H3bomb}
Fandom, T. H. P.~W. 2023.
\newblock Lost episodes.
\newblock \url {https://h3podcast.fandom.com/wiki/Lost_Episodes}.

\bibitem[{Farah(2023)}]{Farah2023}
Farah, H. 2023.
\newblock What is rumble, the video-sharing platform “immune to cancel
  culture”?
\newblock
  https://www.theguardian.com/technology/2023/sep/20/what-is-rumble-the-video-sharing-platform-immune-to-cancel-culture.

\bibitem[{Fisher(2023)}]{Fisher2023}
Fisher, C. 2023.
\newblock Ethan Klein announces another victory in legal battle with Triller.
\newblock
  \url{https://www.dexerto.com/entertainment/ethan-klein-announces-another-victory-in-legal-battle-with-triller-1918577/}.

\bibitem[{gop.gov(2024)}]{gopgov2024}
gop.gov. 2024.
\newblock It is now a fact that Pelosi’s sham January 6th committee was
  designed to be a political witch-hunt.
\newblock https://www.gop.gov/news/documentsingle.aspx?DocumentID=758.

\bibitem[{Grootendorst(2022)}]{grootendorst2022bertopic}
Grootendorst, M. 2022.
\newblock BERTopic: Neural topic modeling with a class-based TF-IDF procedure.
\newblock \emph{arXiv:2203.05794}.

\bibitem[{Grunfeld(2023)}]{Riversidefm23}
Grunfeld, A. 2023.
\newblock Why Has Video Podcasting Become Increasingly Popular?
\newblock \url{https://riverside.fm/blog/why-video-podcasting-is-popular}.

\bibitem[{Guhr et~al.(2021)Guhr, Schumann, Bahrmann, and Böhme}]{guhrfullstop}
Guhr, O.; Schumann, A.-K.; Bahrmann, F.; and Böhme, H.~J. 2021.
\newblock FullStop: Multilingual Deep Models for Punctuation Prediction.

\bibitem[{Hanley, Kumar, and
  Durumeric(2023)}]{hanleyHappenstanceUtilizingSemantic2023}
Hanley, H.~W.; Kumar, D.; and Durumeric, Z. 2023.
\newblock Happenstance: {{Utilizing Semantic Search}} to {{Track Russian State
  Media Narratives}} about the {{Russo-Ukrainian War On Reddit}}.
\newblock In \emph{ICWSM}, volume~17, 327--338.

\bibitem[{Horta~Ribeiro et~al.(2021)Horta~Ribeiro, Jhaver, Zannettou,
  Blackburn, Stringhini, De~Cristofaro, and West}]{horta2021platform}
Horta~Ribeiro, M.; Jhaver, S.; Zannettou, S.; Blackburn, J.; Stringhini, G.;
  De~Cristofaro, E.; and West, R. 2021.
\newblock Do platform migrations compromise content moderation? evidence from
  r/the\_donald and r/incels.
\newblock \emph{Proceedings of the ACM on Human-Computer Interaction},
  5(CSCW2): 1--24.

\bibitem[{Jhaver et~al.(2021)Jhaver, Boylston, Yang, and
  Bruckman}]{jhaver2021evaluating}
Jhaver, S.; Boylston, C.; Yang, D.; and Bruckman, A. 2021.
\newblock Evaluating the effectiveness of deplatforming as a moderation
  strategy on Twitter.
\newblock \emph{Proceedings of the ACM on Human-Computer Interaction},
  5(CSCW2): 1--30.

\bibitem[{Kim(2004)}]{kim2004automatic}
Kim, J. 2004.
\newblock \emph{Automatic detection of sentence boundaries, disfluencies, and
  conversational fillers in spontaneous speech}.
\newblock Ph.D. thesis, Citeseer.

\bibitem[{Kim, Lee, and Park(2016)}]{kimDelineatingComplexUse2016}
Kim, J.; Lee, Y.-O.; and Park, H.~W. 2016.
\newblock Delineating the Complex Use of a Political Podcast in {{South Korea}}
  by Hybrid Web Indicators: {{The}} Case of the {{Nakkomsu Twitter}} Network.
\newblock \emph{TFSC}, 110: 42--50.

\bibitem[{Kim, Kim, and Wang(2016)}]{kimSelectiveExposurePodcast2016}
Kim, Y.; Kim, Y.; and Wang, Y. 2016.
\newblock Selective Exposure to Podcast and Political Participation: The
  Mediating Role of Emotions.
\newblock \emph{IJMC}, 14(2): 133--148.

\bibitem[{Klee(2023)}]{Klee2023}
Klee, M. 2023.
\newblock Rumble is down. will Russell Brand’s allegations knock it out?
\newblock
  \url{https://www.rollingstone.com/culture/culture-features/russell-brand-allegations-rumble-1234851624/}.

\bibitem[{Klein(2023)}]{fasterwhisper}
Klein, G. 2023.
\newblock faster-whisper.
\newblock \url{https://github.com/guillaumekln/faster-whisper}.

\bibitem[{Koenecke et~al.(2024)Koenecke, Choi, Mei, Schellmann, and
  Sloane}]{koenecke2024careless}
Koenecke, A.; Choi, A. S.~G.; Mei, K.; Schellmann, H.; and Sloane, M. 2024.
\newblock Careless Whisper: Speech-to-Text Hallucination Harms.
\newblock \emph{ArXiv:2402.08021}.

\bibitem[{Lai et~al.(2022)Lai, Brown, Bisbee, Tucker, Nagler, and
  Bonneau}]{lai2022estimating}
Lai, A.; Brown, M.~A.; Bisbee, J.; Tucker, J.~A.; Nagler, J.; and Bonneau, R.
  2022.
\newblock Estimating the ideology of political youtube videos.
\newblock \emph{Political Analysis}, 1--16.

\bibitem[{Lapin(2022)}]{Lapin}
Lapin, A. 2022.
\newblock
  \url{https://www.timesofisrael.com/prominent-youtube-personality-suspended-for-holocaust-joke-about-ben-shapiro/}.

\bibitem[{Lee(2021)}]{leeNewsPodcastUsage2021a}
Lee, C. 2021.
\newblock News {{Podcast Usage}} in {{Promoting Political Participation}} in
  {{Korea}}.
\newblock \emph{AJMMC}, 7(2): 107--120.

\bibitem[{Ling et~al.(2021)Ling, AbuHilal, Blackburn, De~Cristofaro, Zannettou,
  and Stringhini}]{ling2021dissecting}
Ling, C.; AbuHilal, I.; Blackburn, J.; De~Cristofaro, E.; Zannettou, S.; and
  Stringhini, G. 2021.
\newblock Dissecting the meme magic: Understanding indicators of virality in
  image memes.
\newblock \emph{CSCW}, 5(CSCW1): 1--24.

\bibitem[{MacDougall(2011)}]{macdougallPodcastingPoliticalLife2011}
MacDougall, R.~C. 2011.
\newblock Podcasting and {{Political Life}}.
\newblock \emph{American Behavioral Scientist}, 55(6): 714--732.

\bibitem[{Mak(2021)}]{Mak2021}
Mak, A. 2021.
\newblock Gab is furious that Donald Trump signed up for another right-wing
  social network.
\newblock
  \url{https://slate.com/technology/2021/06/donald-trump-rally-rumble-gab-parler.html}.

\bibitem[{Maruf and Stelter(2022)}]{MarufStelter}
Maruf, R.; and Stelter, B. 2022.
\newblock
  \url{https://www.cnn.com/2022/02/05/media/joe-rogan-racial-slur-apology-india-arie/index.html}.

\bibitem[{Mayer(2023)}]{Mayer2023b}
Mayer, E. 2023.
\newblock YouTube’s growth as a podcast power player revealed in Cumulus
  Media and signal hill insights’ podcast download – fall 2023 report.
\newblock
  \url{https://www.westwoodone.com/blog/2023/12/04/youtubes-growth-as-a-podcast-power-player-revealed-in-cumulus-media-and-signal-hill-insights-podcast-download-fall-2023-report/}.

\bibitem[{McCarthy(2022)}]{McCarthy2022}
McCarthy, J.~D. 2022.
\newblock Are Antifa and black lives matter related? - The Washington Post.
\newblock
  \url{https://www.washingtonpost.com/politics/2022/02/08/antifa-blm-extremism-violence/}.

\bibitem[{McCluskey(2022)}]{McCluskey2022}
McCluskey, M. 2022.
\newblock Rumble offers Joe Rogan \$100 million to switch platforms.
\newblock \url{https://time.com/6145835/joe-rogan-rumble-podcast-offer/}.

\bibitem[{McInnes, Healy, and Astels(2017)}]{mcinnes2017hdbscan}
McInnes, L.; Healy, J.; and Astels, S. 2017.
\newblock hdbscan: Hierarchical density based clustering.
\newblock \emph{J. Open Source Softw.}, 2(11): 205.

\bibitem[{McInnes, Healy, and Melville(2018)}]{mcinnes2018umap}
McInnes, L.; Healy, J.; and Melville, J. 2018.
\newblock Umap: Uniform manifold approximation and projection for dimension
  reduction.
\newblock \emph{arXiv:1802.03426}.

\bibitem[{McNeil-Willson(2022)}]{mcneil2022understanding}
McNeil-Willson, R. 2022.
\newblock Understanding the\# plandemic: Core framings on Twitter and what this
  tells us about countering online far right COVID-19 conspiracies.
\newblock \emph{First Monday}.

\bibitem[{Mikolov et~al.(2013)Mikolov, Sutskever, Chen, Corrado, and
  Dean}]{mikolov2013distributed}
Mikolov, T.; Sutskever, I.; Chen, K.; Corrado, G.~S.; and Dean, J. 2013.
\newblock Distributed representations of words and phrases and their
  compositionality.
\newblock \emph{NIPS}, 26.

\bibitem[{Milbauer, Mathew, and Evans(2021)}]{milbauer2021aligning}
Milbauer, J.; Mathew, A.; and Evans, J. 2021.
\newblock Aligning multidimensional worldviews and discovering ideological
  differences.
\newblock In \emph{EMNLP}.

\bibitem[{Mittal et~al.(2024)Mittal, Murthy, Kumar, and
  Bhat}]{mittal2024towards}
Mittal, A.; Murthy, R.; Kumar, V.; and Bhat, R. 2024.
\newblock Towards understanding and mitigating the hallucinations in NLP and
  Speech.
\newblock In \emph{11th ACM IKDD CODS and 29th COMAD}, 489--492.

\bibitem[{NIST(2003)}]{NIST2003}
NIST. 2003.
\newblock The Rich Transcription Fall 2003 (RT-03F) Evaluation Plan.
\newblock
  \url{http://www.nist.gov/speech/tests/rt/rt2003/fall/docs/rt03-fall-eval-plan-v9.pdf}.

\bibitem[{OpenAI(2022)}]{OpenAIWhisper}
OpenAI. 2022.
\newblock Whisper.
\newblock \url{https://github.com/openai/whisper}.

\bibitem[{OpenNMT(2019)}]{ctranslate2}
OpenNMT. 2019.
\newblock CTranslate2.
\newblock \url{https://github.com/OpenNMT/CTranslate2}.

\bibitem[{Patel et~al.(2024)Patel, Paudel, De~Cristofaro, Stringhini, and
  Blackburn}]{patel2024idrama}
Patel, J.; Paudel, P.; De~Cristofaro, E.; Stringhini, G.; and Blackburn, J.
  2024.
\newblock iDRAMA-Scored-2024: A Dataset of the Scored Social Media Platform
  from 2020 to 2023.
\newblock In \emph{ICWSM}, volume~18, 2014--2024.

\bibitem[{Pedregosa et~al.(2011)Pedregosa, Varoquaux, Gramfort, Michel,
  Thirion, Grisel, Blondel, Prettenhofer, Weiss, Dubourg
  et~al.}]{pedregosa2011scikit}
Pedregosa, F.; Varoquaux, G.; Gramfort, A.; Michel, V.; Thirion, B.; Grisel,
  O.; Blondel, M.; Prettenhofer, P.; Weiss, R.; Dubourg, V.; et~al. 2011.
\newblock Scikit-learn: Machine learning in Python.
\newblock \emph{JMLR}, 12: 2825--2830.

\bibitem[{Pew(2006)}]{Pew2006}
Pew. 2006.
\newblock What is podcasting?
\newblock
  \url{https://www.pewresearch.org/journalism/2006/07/19/what-is-podcasting/}.

\bibitem[{Podcast(2024)}]{weinstein_2024}
Podcast, H. 2024.
\newblock \url{https://www.doesryankavanaughlooklikeharveyweinstein.com/}.

\bibitem[{Pramod(2021)}]{Pramod2021}
Pramod, N. 2021.
\newblock Rumble is experiencing massive growth as people ditch Big Tech.
\newblock
  \url{https://reclaimthenet.org/rumble-is-experiencing-massive-growth}.

\bibitem[{Putri et~al.(2022)Putri, Ermayda, Firmansyah, Sulistyawati, and
  Mariyah}]{putri2022radio}
Putri, D.~M.; Ermayda, R.~Z.; Firmansyah, R.; Sulistyawati, S.~N.; and Mariyah,
  S. 2022.
\newblock Radio Audit: A Visual Podcast Series as The Learning Media to
  Increase Accounting Students Engagement.
\newblock \emph{Ekuitas: Jurnal Pendidikan Ekonomi}, 10(2): 219--227.

\bibitem[{Radford et~al.(2021)Radford, Kim, Hallacy, Ramesh, Goh, Agarwal,
  Sastry, Askell, Mishkin, Clark et~al.}]{radford2021learning}
Radford, A.; Kim, J.~W.; Hallacy, C.; Ramesh, A.; Goh, G.; Agarwal, S.; Sastry,
  G.; Askell, A.; Mishkin, P.; Clark, J.; et~al. 2021.
\newblock Learning transferable visual models from natural language
  supervision.
\newblock In \emph{ICML}, 8748--8763. PMLR.

\bibitem[{Rae(2023)}]{raePodcastsPoliticalListening2023}
Rae, M. 2023.
\newblock Podcasts and Political Listening: Sound, Voice and Intimacy in the
  {{Joe Rogan Experience}}.
\newblock \emph{Continuum}, 37(2): 182--193.

\bibitem[{Rainey(2023)}]{Rainey2023}
Rainey, C. 2023.
\newblock Is rumble right-wing? CEO argues it has many left-leaning ...
\newblock
  \url{https://www.fastcompany.com/90982735/rumble-video-site-youtube-alternative-left-leaning-users}.

\bibitem[{Rajic(2013)}]{rajic2013educational}
Rajic, S. 2013.
\newblock Educational use of podcast.
\newblock In \emph{The Fourth International Conference on e-Learning}, 90--94.

\bibitem[{Robertson(2019)}]{Robertson2019}
Robertson, J. 2019.
\newblock How podcasts went from Unlistenable to unmissable.
\newblock \url{https://www.bbc.com/news/business-49279177}.

\bibitem[{Rosman et~al.(2022)Rosman, Sisario, Isaac, and
  Satariano}]{Rosman2022}
Rosman, K.; Sisario, B.; Isaac, M.; and Satariano, A. 2022.
\newblock Spotify Bet Big on Joe Rogan. it got more than it counted on.
\newblock
  \url{https://www.nytimes.com/2022/02/17/arts/music/spotify-joe-rogan-misinformation.html}.

\bibitem[{Rumble(2023)}]{rumblepodcasts}
Rumble. 2023.
\newblock Podcasts.
\newblock
  \url{https://web.archive.org/web/20230131060622/https://rumble.com/category/podcasts}.

\bibitem[{Sayogie et~al.(2023)Sayogie, Farkhan, Julian, Al~Hakim, Wiralaksana
  et~al.}]{sayogie2023patriarchal}
Sayogie, F.; Farkhan, M.; Julian, H.~P.; Al~Hakim, H. S.~F.; Wiralaksana,
  M.~G.; et~al. 2023.
\newblock Patriarchal Ideology, Andrew Tate, and Rumble's Podcasts.
\newblock \emph{3L: Southeast Asian Journal of English Language Studies},
  29(2).

\bibitem[{Shearer et~al.(2023)Shearer, Liedke, Matsa, Lipka, and
  Jurkowitz}]{Pew2023}
Shearer, E.; Liedke, J.; Matsa, K.~E.; Lipka, M.; and Jurkowitz, M. 2023.
\newblock Podcasts as a Source of News and Information.
\newblock \emph{Pew Research Center}, 16.

\bibitem[{Sherman(2024)}]{Sherman2024}
Sherman, C. 2024.
\newblock Many Republicans support abortion. are they switching parties because
  of it?
\newblock
  \url{https://www.theguardian.com/world/2024/jan/13/abortion-republican-voters-presidential-election}.

\bibitem[{Spangler(2022)}]{Spangler2022}
Spangler, T. 2022.
\newblock Joe Rogan rejects 100 million podcast deal offer from right-wing
  video site rumble.
\newblock \url
  {https://variety.com/2022/digital/news/joe-rogan-rejects-rumble-podcast-deal-1235175990/}.

\bibitem[{Sterne et~al.(2008)Sterne, Morris, Baker, and
  Freire}]{sternePoliticsPodcasting2008}
Sterne, J.; Morris, J.; Baker, M.~B.; and Freire, A.~M. 2008.
\newblock The {{Politics}} of {{Podcasting}}.
\newblock \emph{Fibreculture}.

\bibitem[{Stocking et~al.(2022)Stocking, Mitchell, Matsa, Widjaya, Jurkowitz,
  Ghosh, Smith, Naseer, and St~Aubin}]{stocking2022role}
Stocking, G.; Mitchell, A.; Matsa, K.~E.; Widjaya, R.; Jurkowitz, M.; Ghosh,
  S.; Smith, A.; Naseer, S.; and St~Aubin, C. 2022.
\newblock The role of alternative social media in the news and information
  environment.
\newblock \emph{Pew Research Center}.

\bibitem[{TateSpeech(2024)}]{TateSpeechRumble}
TateSpeech. 2024.
\newblock {TateSpeech - Rumble}.
\newblock \url{https://rumble.com/c/TateSpeech}.
\newblock Accessed: 2024-01-12.

\bibitem[{Team(2021)}]{SileroVAD}
Team, S. 2021.
\newblock Silero VAD: pre-trained enterprise-grade Voice Activity Detector
  (VAD), Number Detector and Language Classifier.
\newblock \url{https://github.com/snakers4/silero-vad}.

\bibitem[{Wakabayashi, Alba, and Tracy(2020)}]{Wakabayashi2020}
Wakabayashi, D.; Alba, D.; and Tracy, M. 2020.
\newblock Bill Gates, at odds with trump on virus, becomes a right-wing target.
\newblock
  \url{https://www.nytimes.com/2020/04/17/technology/bill-gates-virus-conspiracy-theories.html}.

\bibitem[{Wilson(2022)}]{Wilson2022}
Wilson, J. 2022.
\newblock The downfall of Andrew Tate and its implications.
\newblock
  \url{https://www.forbes.com/sites/joshwilson/2022/08/30/the-downfall-of-andrew-tate-and-its-implications}.

\bibitem[{Yang, Jang, and Yu(2023)}]{yang2023analyzing}
Yang, J.; Jang, H.; and Yu, K. 2023.
\newblock Analyzing Geographic Questions Using Embedding-based Topic Modeling.
\newblock \emph{ISPRS International Journal of Geo-Information}, 12(2): 52.

\bibitem[{YouTube(2023)}]{YouTubepopular}
YouTube. 2023.
\newblock youtube-popularcreators.
\newblock \url{https://www.youtube.com/podcasts/popularcreators}.

\bibitem[{ytdl(2006)}]{youtubedl}
ytdl. 2006.
\newblock youtube-dl.
\newblock \url{https://github.com/ytdl-org/youtube-dl}.

\bibitem[{Zannettou et~al.(2018)Zannettou, Caulfield, Blackburn, De~Cristofaro,
  Sirivianos, Stringhini, and Suarez-Tangil}]{zannettou2018origins}
Zannettou, S.; Caulfield, T.; Blackburn, J.; De~Cristofaro, E.; Sirivianos, M.;
  Stringhini, G.; and Suarez-Tangil, G. 2018.
\newblock On the origins of memes by means of fringe web communities.
\newblock In \emph{IMC}, 188--202.

\end{thebibliography}

\appendix

\section{A}

\descr{Language verification for podcasts.}
In addition to our initial step of excluding non-English channels and playlists, following previous work~\cite{clifton100000Podcasts2020}, we run language detection on podcast video descriptions.
For this purpose, we use langdetect library~\cite{langdetect}, which is a Python implementation of Google's language detection library in Java.
We also remove URLs from video descriptions before running language detection.
During a manual inspection of videos flagged as non-English, we observe that these videos have short descriptions (e.g., social media platforms and their URLs) that could cause mislabeling their languages. 
Consequently, we conduct a manual inspection of these videos and videos with no description, and exclude ``Monarky'' channel from Rumble, due to its content being in a language other than English.

\descr{Speech-to-Text transcription.}
For the transcription of podcast videos, we use faster-whisper~\cite{fasterwhisper}, a reimplementation of OpenAI's Whisper~\cite{OpenAIWhisper} via CTranslate2~\cite{ctranslate2}, in conjunction with Silero's Voice Activity Detection~\cite{SileroVAD}.
This combination is particularly effective in handling challenges (e.g., long pauses and background music) present in many videos in our dataset.
We use the large-v2 model of Whisper in our analysis and use English as the language parameter.
In total, we spend 658 hours (27 days) with NVIDIA A100 GPU with 80GB of Memory to generate their speech-to-text transcriptions.

\descr{Postprocessing.}
We implement three postprocessing steps.
To refine our analysis, we remove English stop words from the topic keywords using Scikit-learn's CountVectorizer function~\cite{pedregosa2011scikit}.
Next, we exclude topics that comprise fewer than 5 keywords. 
This decision is based on our observation from manually inspecting the top 100 most popular topics, which indicates that topics with few keywords predominantly consist of generic sentences that are mostly identical (e.g., ``Ok.'').
Subsequently, we filter out topics characterized by conversational fillers and backchannels, e.g., ``hmm,'' ``yeah,'' ``oh,'' ``uh,'' ``so,'' and ``well,'' if these appear among a topic's top five keywords.
For this purpose, we use a keyword list derived from previous work~\cite{kim2004automatic}, which is constructed based on annotated conversational speech data from the Linguistic Data Consortium and standard scoring tools~\cite{NIST2003}.
This step is crucial as our primary goal is to enhance the interpretability of our results. 
Nonetheless, we perform no additional postprocessing due to the intrinsic characteristics of podcast content, which may include casual or mundane discussions.
We treat the remaining generic topics as indicative of everyday conversation, providing a richer, more nuanced understanding of our findings.

\begin{figure}[ht]
\centering
\includegraphics[width=0.98\linewidth]{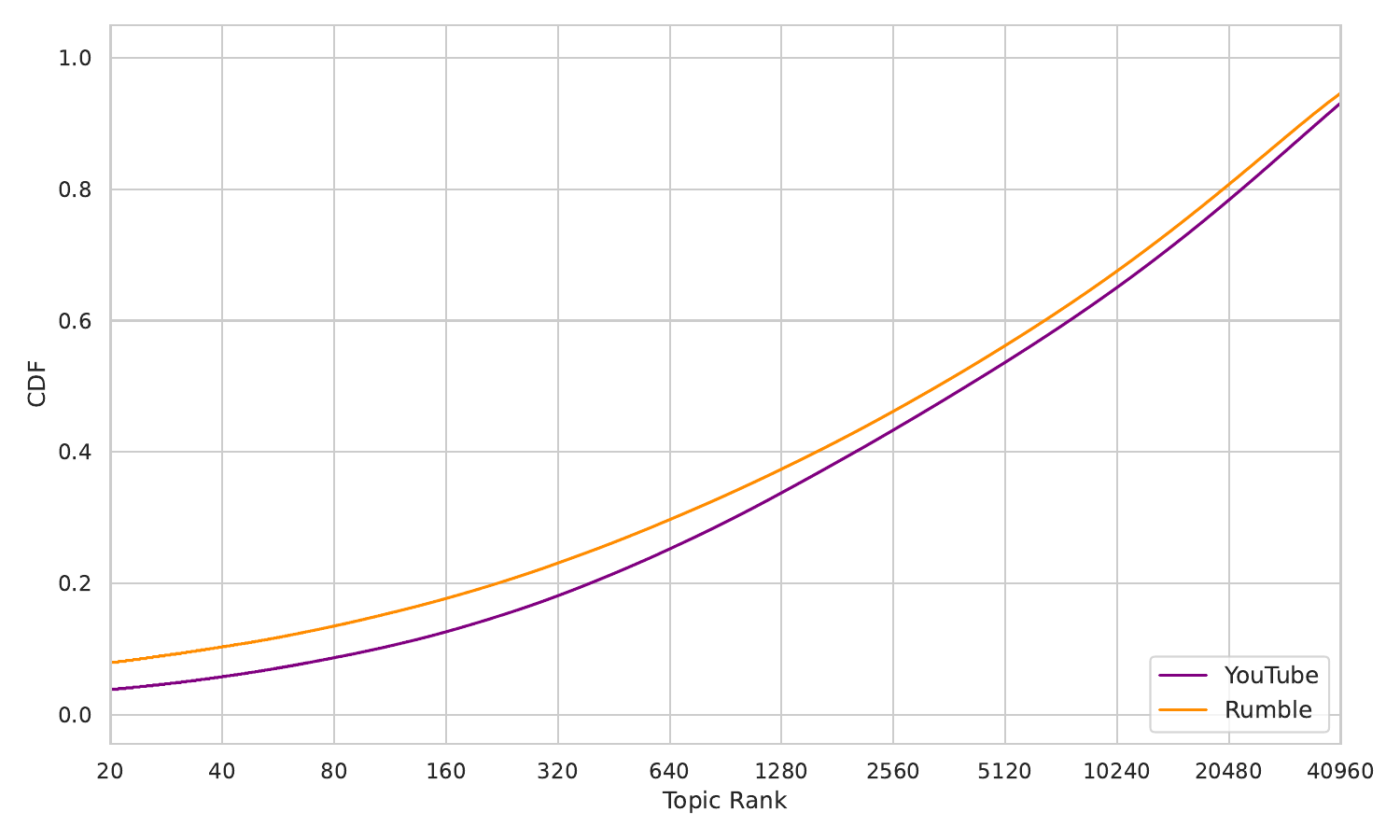}
\caption{CDF of the proportion of sentences covered cumulatively at each topic rank in YouTube and Rumble podcast videos. Topic ranks start at 20 and increase exponentially.}
\label{fig:cdf_topic_ranks}
\end{figure}

\begin{figure*}[h]
\centering
\includegraphics[width=0.8\linewidth]{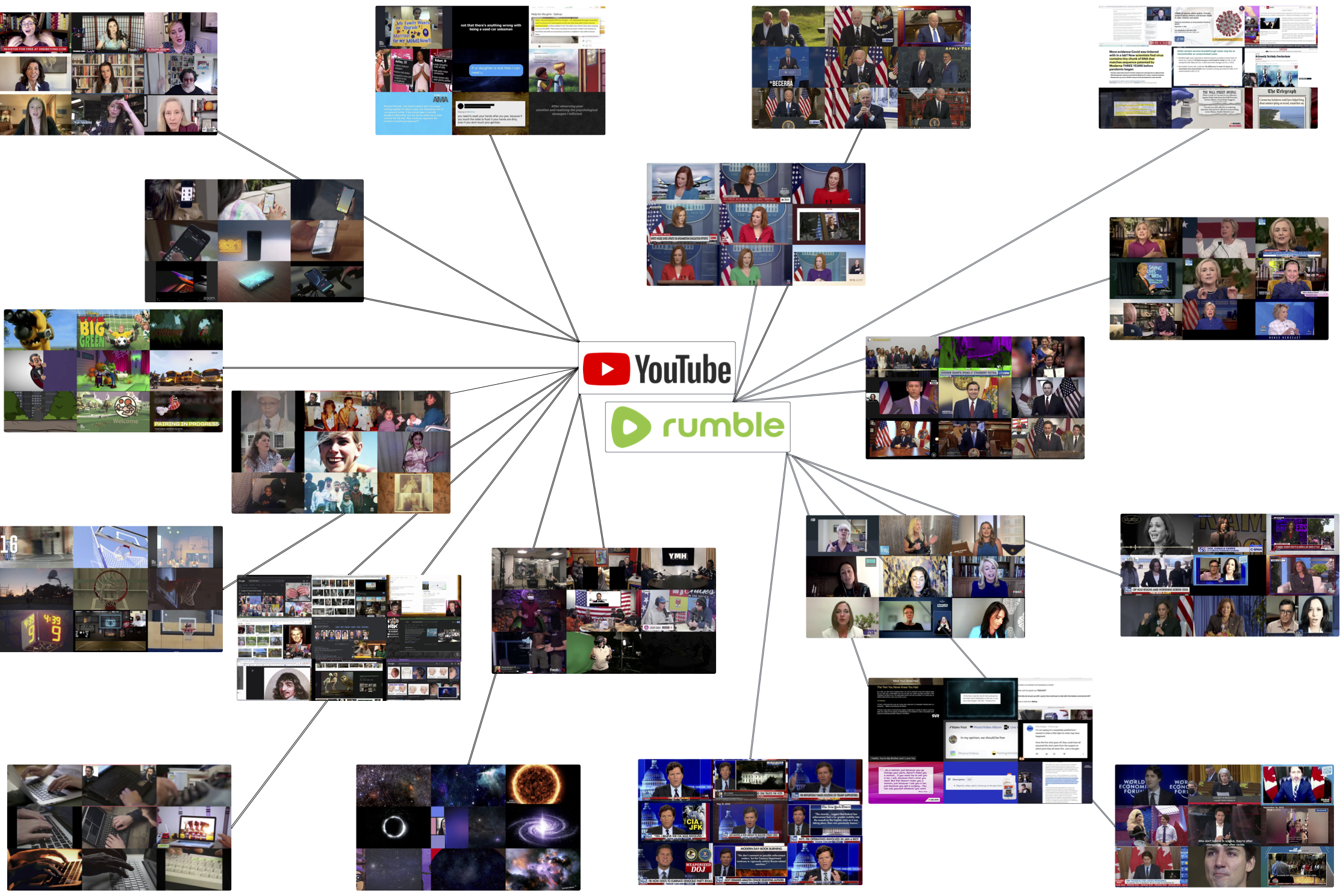}
\caption{Comparison of visual topics between Youtube and Rumble, extracted through clustering, showing top-10 clusters for each platform (Refer to Table~\ref{tab:comparison_top_images}).}
\label{fig:top10_cluster}
\end{figure*}

\section{B}

\descr{Misalignment analysis.}
To further solidify our findings for RQ1, we analyze the differences in word usages between \dsrumble and \dsyoutube.
To do this, we leverage the methodlogy proposed by Milbauer et al.~\cite{milbauer2021aligning}, which trains word2vec models for each community, and aligns their words using a linear translation function MultiCCA~\cite{ammar2016massively}.
If a community's word projection does not match the same word in another community, we consider these words are \emph{misaligned}.
This way, by identifying misaligned word pairs with political meanings, e.g., Democrat's usage of ``Republican'' and Republican's usage of ``Democrat, '' we can have an understanding of a community's political positioning.

\descr{Training.}
We follow the preprocessing steps proposed by Milbauer et al., where we tokenize each sentence, remove hyperlinks, and lowercase all characters.
Next, we train  Word2Vec skip-gram models~\cite{mikolov2013distributed} for \dsyoutube and \dsrumble using 100 dimensions and a maximum vocabulary of 30,000 words.  
We anchor top 5K common words of these datasets and translate them using MultiCCA.

\descr{Results.}
Table~\ref{tab:alignment} presents identified misaligning word pairs between \dsyoutube and \dsrumble, along with their cosine similarities. 
Similar to our previous example, we find many misaligning word pairs in the context of ``Democrats vs Republicans.''
This is evident from Republicans \& Democrats, Democrat \& Republican, Dems \& Democrats, Leftists \& Right-wingers, Hillary Clintons \& Trumps, Progressive \& Conservative, and Pro-Trump \& Anti-Trump word pairs.

Additionally, we identify Pro-choice \& Pro-life, Antifa \& BLM pairs, and Witch Hunt \& January 6th pairs, which further indicate that \dsrumble aligns with general right-wing narratives on these topics~\cite{McCarthy2022, gopgov2024, Sherman2024}. 
Overall, these results further solidifies our findings from RQ1, demonstrating that Rumble's podcast content exhibits a pronounced right-wing bias, a trend that remains evident even when compared to YouTube's predominantly apolitical content.

\begin{figure}[ht]
  \centering
  \includegraphics[width=0.98\linewidth]{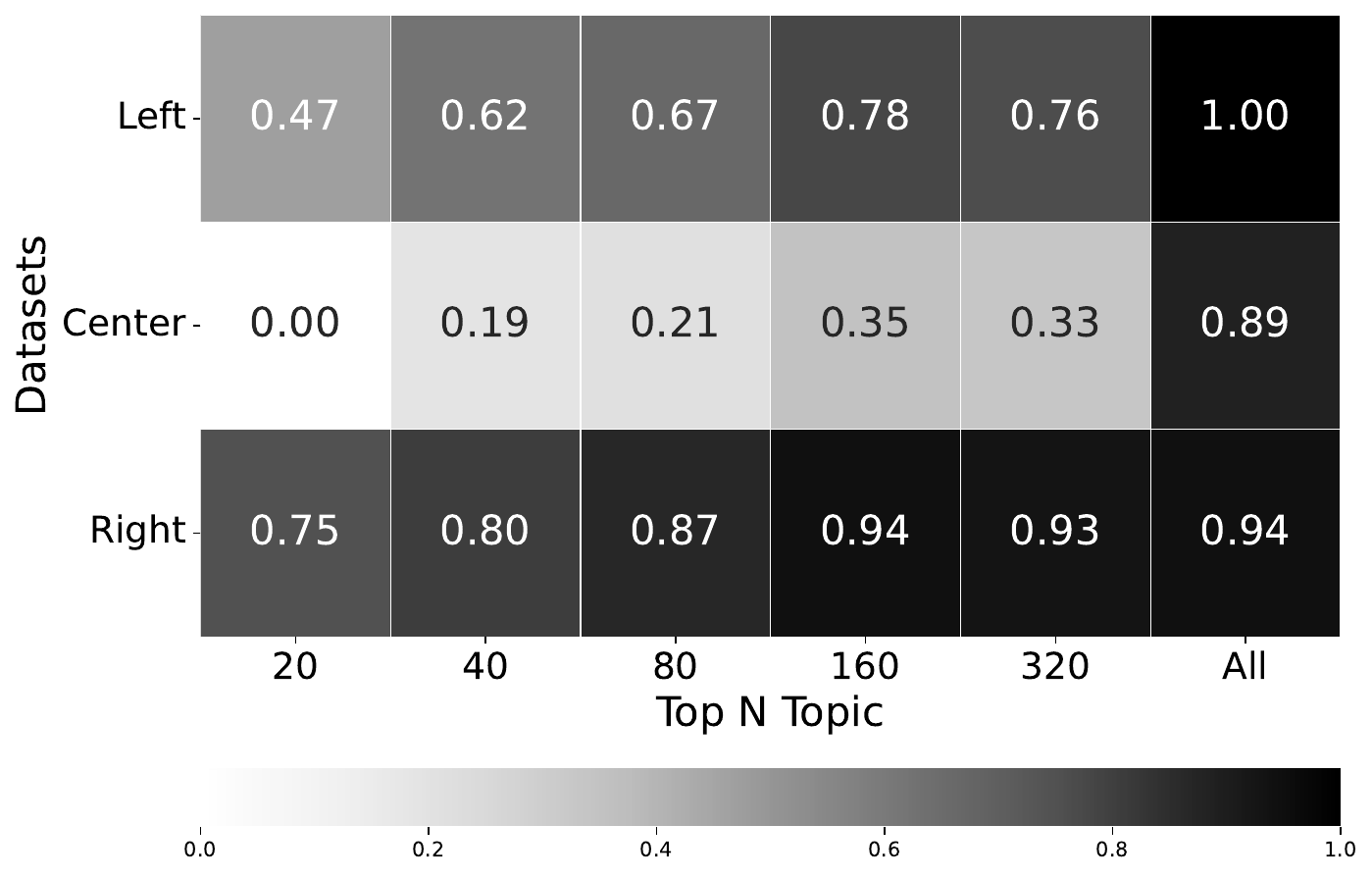}
  \caption{Heatmap illustrating the normalized cosine similarities among the top N topic centroids of Rumble versus left-wing, center, and right-wing podcasts. Darker shades denote greater semantic similarity.}
\label{fig:rumble_semantic_normalized}
\end{figure}

\begin{table}[t!]
\centering
\begin{tabular}{@{}llr@{}}
\toprule
\textbf{Rumble}         & \textbf{YouTube}     & \textbf{Alignment} \\ \midrule
Republicans             & Democrats            & 0.8787             \\
Democrat                & Republican           & 0.7717             \\
Dems                    & Democrats            & 0.6986             \\
Leftists                & Right-wingers        & 0.6231             \\
Hillary Clintons        & Trumps               & 0.5761             \\
Pro-choice              & Pro-life             & 0.5560             \\
Progressive             & Conservative         & 0.5190             \\
Antifa                  & BLM                  & 0.5048             \\
Pro-Trump               & Anti-Trump           & 0.4732             \\
Witch Hunt              & January 6th          & 0.4571             \\ \bottomrule
\end{tabular}
\caption{Identified misaligning word pairs between popular podcast channels of  YouTube and Rumble.}
\label{tab:alignment}
\end{table}

\section{C}

\descr{Clustering.}
We leverage OpenAI's CLIP~\cite{radford2021learning} to generate embeddings, using its top performing model, \emph{ViT-L/14@336px}.
Our clustering approach is inspired by techniques used in BERTopic~\cite{grootendorst2022bertopic} and Top2Vec~\cite{angelov2020top2vec}. 
This methodology first reduces the dimensionality of these embeddings with UMAP~\cite{mcinnes2018umap}.
Subsequently, we input these reduced-dimension embeddings into HDBSCAN~\cite{mcinnes2017hdbscan}, an algorithm that excels in generating dense clusters without the need for predefining cluster sizes.
This flexibility allows us to explore thematic topics organically, without the constraint of limiting the visual clusters to a specific number.

\begin{table}[h]
  \centering
  \scriptsize

  \begin{tabular}{lrrrr}
  \toprule
              Channel Name &  \# Sentences & \# Podcasts &    Avg. views \\
  \midrule
                
          The Dan Bongino Show &        42 &          31 &               198,535 \\
          The Charlie Kirk Show &            41  &          31 &                 27,331 \\
          Ben Shapiro &            33 &          16 &                 8,925 \\
          Rekieta Law &            33 &          9 &                 30,361 \\
          phetasy  &            32 &           21 &              2,425 \\
          TateSpeech by Andrew Tate &            31 &          2 &                3,775,000 \\
          vivafrei  &            25 &          14 &                28,579 \\
          Matt Walsh &            21 &          11 &                7,912 \\
          Steven Crowder &            20 &          15 &              237,480 \\
          Dinesh D'Souza &            16 &          14 &                14,494 \\
          TimcastIRL  &            16 &          11 &                8,280 \\
          AMERICA First with Sebastian Gorka  &            15 &          11 &                11,880 \\
          The Dershow &            15 &          11 &                 9,530 \\
          The Rubin Report &            13 &          8 &               48,412 \\
          Liz Wheeler  &            12 &          12 &                3,674 \\
          Diamond and Silk &            11 &           5 &                 16,702 \\
          Lara Trump &            11 &           6 &                12,490 \\
          Mikhaila Peterson &           11 &          4 &                13,437 \\
          The Ron Paul Liberty Report &            11  &           10 &               31,305 \\
          Redacted News &            8 &          5 &                  31,780 \\
  \bottomrule
  \end{tabular}
  \caption{Top 20 channels on Rumble related to the topic of cancellation. We present data, including the total number of sentences, the overall number of podcast videos, and the average views per podcast video for each channel within this topic.}
\label{tab:cancelled_info}
\end{table}

\end{document}